\documentclass[preprint]{aastex}

\begin{document}

\title{A CHANDRA STUDY OF SGR~A EAST: A SUPERNOVA REMNANT REGULATING THE
   ACTIVITY OF OUR GALACTIC CENTER?}

\author{Y.\ Maeda,\altaffilmark{1} 
F.\ K.\ Baganoff,\altaffilmark{2} 
E.\ D.\ Feigelson,\altaffilmark{1} 
M.\ Morris,\altaffilmark{3} 
M.\ W.\ Bautz,\altaffilmark{2} 
W.\ N.\ Brandt,\altaffilmark{1} 
D.\ N.\ Burrows,\altaffilmark{1} 
J.\ P.\ Doty,\altaffilmark{2}
G.\ P.\ Garmire,\altaffilmark{1} 
S.\ H.\ Pravdo,\altaffilmark{4} 
G.\ R.\ Ricker,\altaffilmark{2} 
L.\ K.\ Townsley\altaffilmark{1} 
}

\altaffiltext{1}{Department of Astronomy and Astrophysics, 525 Davey
Laboratory, The Pennsylvania State University, University Park, PA
16802-6305, U.S.A.}

\altaffiltext{2}{Massachusetts Institute of Technology, Center for
Space Research, 77 Massachusetts Avenue, Cambridge, MA 02139-4307,
U.S.A.}

\altaffiltext{3}{Division of Astronomy, Box 951562, UCLA, Los Angeles,
CA 90095-1562, U.S.A.}

\altaffiltext{4}{Jet Propulsion Laboratory, MS 306-438, 4800 Oak Grove
Drive, Pasadena, CA 91109, U.S.A.}

\slugcomment{Revised Version 1}

\shorttitle{Chandra ACIS Imaging Spectroscopy of Sgr~A East}

\shortauthors{Y.~Maeda et al.}

\begin{abstract}

We report on the X-ray emission from the shell-like, non-thermal radio
source Sgr~A East (SNR~000.0$+$00.0) located in the inner few parsecs
of the Galaxy based on observations made with the ACIS detector on
board the {\it Chandra X-ray Observatory}.  This is the first time
Sgr~A East has been clearly resolved from other complex structures in
the region. The X-ray emitting region is concentrated within the
central $\simeq 2$ pc of the larger radio shell. The spectrum shows
strong K$\alpha$ lines from highly ionized ions of S, Ar, Ca, and Fe.
A simple isothermal plasma model gives electron temperature $\sim
2$~keV, absorption column $\sim 1 \times 10^{23}$~H~cm$^{-2}$,
luminosity $\sim 8 \times 10^{34}$~ergs~s$^{-1}$ in the 2--10 keV
band, and gas mass $\sim 2\eta^{\frac{1}{2}}$ M$_{\odot}$ with a
filling factor $\eta$. The plasma appears to be rich in heavy
elements, over-abundant by roughly a factor of four with respect to
solar abundances, and shows a spatial gradient of elemental abundance:
the spatial distribution of iron is more compact than that of the
lighter elements.

The gas mass and elemental abundance of the X-ray emission support the
long-standing hypothesis that Sgr~A East is a supernova remnant (SNR),
maybe produced by the Type~II supernova explosion of a massive star
with a main-sequence mass of 13--20 M$_\odot$. The combination of the
radio and X-ray morphologies classifies Sgr~A East as a new metal-rich
``mixed morphology'' (MM) SNR. The size of the Sgr~A East radio shell
is the smallest of the known MM SNRs, which strongly suggests that the
ejecta have expanded into a very dense interstellar medium. The
ejecta-dominated chemical compositions of the plasma indicate that the
ambient materials should be highly homogeneous.

We thus evaluate a simplified dynamical evolution model where a SNR
was formed about 10,000 years ago and expanded into an ambient medium
with a homogeneous density of $10^3$ cm$^{-3}$. The model roughly
reproduces most of the observed properties in the X-ray and radio
wavelengths. A comparison with the radio observations requires the dense
ambient medium to be ionized, but a luminous X-ray irradiator with an
expected X-ray luminosity of $\sim 10^{40}$ ergs~s$^{-1}$ is not
currently present.  The presence of the ionized gas may be explained
if the massive black hole (MBH) associated with the compact,
non-thermal radio source Sgr~A$^*$ was bright in X-rays about three
hundred years ago, but is presently dim.  It is possible that the
dust/molecular ridge compressed by the forward shock of Sgr~A East hit
Sgr~A* in the past, and the passage of the ridge may have supplied
material to accrete onto the black~hole in the past, and may have
removed material from the black hole vicinity, leading to its present
quiescent state. This may be a specific example of the intimate
relationship between a SNR and massive black~hole accretion activity
in galactic nuclei.

\end{abstract}


\clearpage

\section{INTRODUCTION}

The center of our Galaxy embodies a rich variety of phenomena which
create diverse complex structures that are visible to us over a broad
range of wavelengths. The radio emission from the central few parsecs
of the Galaxy has several components, including a compact non-thermal
source (Sgr~A$^*$) thought to be associated with the central massive
black hole, a spiral-shaped group of thermal gas streams (Sgr~A West)
that are possibly infalling to Sgr~A$^*$, and a 3\arcmin.5 $\times$
2\arcmin.5 shell-like non-thermal structure (Sgr~A East) \citep[][see
also Figure~\ref{figure:1}]{Ekers75}. Sgr~A East surrounds
Sgr~A$^*$/West in projection, but its center is offset by about
$50\arcsec$ (2~pc). A number of arguments suggest that Sgr~A$^*$/West
is physically located very near or possibly embedded within Sgr~A
East. For the latter case, interaction between Sgr~A East and
Sgr~A$^*$/West would be inevitable, so Sgr~A East may be a key for
understanding the activity in the nucleus of our Galaxy (for a recent
review, see Yusef-Zadeh{,} Melia \& Wardle 2000).

Radio studies indicate that Sgr~A East may be interpreted simply as a
supernova remnant \citep[SNR~000.0+00.0;][]{Jones74,Ekers83,Green84}. However,
its location extremely close to the Galactic nucleus and its inferred
energetics have inspired alternative interpretations: for example,
multiple SNRs or the remnant of an extremely energetic explosion due
to tidal disruption of a star by the central massive black hole
\citep{Yusef87, Mezger89, Khokhlov96}. Thus the origin of Sgr~A East
is still an open issue.

The non-thermal radio emission from the Sgr~A East shell, due to
synchrotron radiation from relativistic electrons, has an unusually
high surface brightness for a Galactic SNR and is an outlier in the
$\Sigma-D$ relation \citep{Green84, Case98}. Yusef-Zadeh et~al. (1996)
found a possible Zeeman-split OH maser line arising in the compressed
dust/molecular ridge outside this non-thermal radio shell, and they
inferred a magnetic field strength of 2--4 mG. This strong magnetic
field may cause unusually rapid synchrotron aging. Regardless of what
actually caused the initial explosion, the co-existence of the
relativistic particles and the strong magnetic field indicates Sgr~A
East is a unique particle accelerator in the Galaxy
\citep[e.g.,][]{Yusef00}. The age and the shock velocity of Sgr~A East
are crucial parameters for determining the energy distribution of the
accelerated particles \citep[][and references therein]{Sturner97},
which leads in turn to a better understanding of how Sgr~A East may
contribute cosmic-rays and $\gamma$-ray emission to the Galactic
Center maelstrom.

Sgr~A East appears to be interacting with the $+$50~km~s$^{-1}$
molecular cloud (M$-0.02-0.07$), and it has been suggested by others
that it could have stimulated star formation as evidenced by a chain
of compact HII regions located just to the east in projection
\citep[][see also Figure~\ref{figure:1}]{Goss85, Ho85, Yusef95,
Novak00}. Cotera et~al. (1999) found an infrared star positionally
coincident with one of the HII regions. They estimated that the star
might be in the Wolf-Rayet phase (WN7), so if the HII region was
created by an interaction with the Sgr~A East shell, then Sgr~A East
would have to be over $\sim 10^{5-6}$ yrs old. However, Uchida
et~al. (1998) showed that the shear associated with non-solid Galactic
rotation causes distortion of an expanding bubble in the Galactic
longitudinal direction. The elongated radio structure of Sgr~A East is
naturally reproduced on relatively short timescales of $\sim 10^{4}$
years, and it would be sheared out of existence in $\sim 10^5$ yrs.

In the soft X-ray band, enhanced emission from the vicinity of Sgr~A
East is evident in a {\it ROSAT} image, but it was not studied
\citep{Predehl94}.  In the hard 2--10 keV band, an {\it ASCA} image
showed an oval-shaped region of $2\arcmin \times 3\arcmin$ filling the
Sgr~A radio shell \citep{Koyama96a} with a surface brightness 5 times
that of the surrounding diffuse emission. The absorption column
density and the luminosity from the oval-shaped region were found to
be approximately $7 \times 10^{22}$ cm$^{-2}$ and $10^{36}$ erg
s$^{-1}$. With {\it BeppoSAX} data, Sidoli et~al. (1999) reported that
the spectrum of the emission integrated over the entire Sgr~A region
can be modeled with a multi-temperature thermal plasma or a
combination of non-thermal and thermal emission with prominent
emission lines at 2.4~keV from S~K$\alpha$ and at 6.7~keV from highly
ionized Fe~K$\alpha$ \citep{Sidoli99b}. However, the $\sim 1\arcmin$
spatial resolution of {\it ASCA} and {\it BeppoSAX} could not
establish the detailed X-ray properties of the Sgr~A region.

The Chandra X-ray Observatory ({\it Chandra}) with the Advanced CCD
Imaging Spectrometer (ACIS) detector combines the wide-band
sensitivity and moderate spectral resolution of the {\it ASCA} and
{\it BeppoSAX} satellites with the much higher spatial resolution
(0\farcs5--1\arcsec) of {\it Chandra}'s High-Resolution Mirror
Assembly (HRMA). ACIS clearly resolved Sgr~A East from its complex
environs. Paper~I \cite{Baganoff00} provides an overview of our
findings for the entire 17\arcmin\ ($\approx 40$~pc) ACIS-I field of
view and discusses the X-ray emission from the immediate vicinity of
the MBH at Sgr~A$^*$.  This paper focuses on a detailed analysis of
Sgr~A East. Throughout this paper we adopt a distance of 8.0 kpc to
the Galactic Center \citep{Reid93}.

\section{OBSERVATIONS AND ANALYSIS PROCEDURES}

\subsection{Data Acquisition and Reduction}

The observation of Sgr~A was carried out early in the {\it Chandra}
mission on 1999 September 21 ($\rm ObsID = 242$) over a period of
51.1~ks using the ACIS-I array of four abutted, frontside-illuminated
CCDs.  The satellite and instrument are described by Weisskopf et al.\
(1996) and Garmire et al.\ (2000) respectively, and details about ACIS
can be found at {\it http://www.astro.psu.edu/xray/axaf}. The
telemetry limit of the satellite was exceeded during $\sim21$\% of the
observation due to high background caused by the impact of energetic
solar particles on the CCDs.  Exposure frames were dropped during
telemetry saturation, causing irrecoverable loss of information.  The
effective exposure time was 40.3~ks.  The Sgr~A complex was imaged
near the center of the ACIS-I array of four $1024 \times 1024$-pixel
CCDs, each with $0\farcs5 \times 0\farcs5$ pixels and a field of view
of $8\farcm4 \times 8\farcm4$.  Data acquisition with ACIS was made in
timed-exposure (TE) very-faint (VF) mode with the chips read out every
3.24~s.  The focal plane temperature was about $-110$ $^\circ$C.

Individual events were pre-processed on-board using a lower event
threshold of 38 ADU and a ``split'' threshold of 13 ADU. Events with
ACIS flight grades of 24, 66, 107, 214, or 255, and those occurring in
known bad pixels and columns, were removed on-board to reduce
telemetry.  The celestial coordinates of events were determined during
ground-based processing based on guide-star aspect solutions, improved
by alignment to astrometric standard stars as described in Paper~I.

Early in the mission, the ACIS frontside-illuminated CCDs suffered a
significant increase in parallel charge transfer inefficiency (CTI)
due to radiation damage acquired during satellite perigee passages
through the terrestrial radiation belts \citep{Prigozhin00}.  The CTI
causes a progressive row-dependent decrease in the detection
efficiency and energy gain accompanied by a degradation of the energy
resolution.  Sgr~A East was imaged near the top of amplifier 3 on chip
I3, a location known to be heavily affected by CTI.  To mitigate the
energy bias and recover from grade migration caused by CTI, we applied
the software corrector described by Townsley et al.\ (2000).

The conversion from event ADU to photon energy used in our spectral
analysis uses a response matrix based on a nearly contemporaneous
observation of reference radioactive and fluorescent emission lines
from an on-board calibration source ($\rm ObsID = 1310$), which were
processed with the same CTI corrector. From analysis of the
instrumental Ni-K$\alpha$ (7.5~keV) and Au-L$\alpha$ (9.7~keV) lines
arising from particle bombardment of satellite metals, we found the
energies are probably underestimated by $\sim 100$~eV at $E \geq 7$
keV. Thus we removed an $\sim$2\% bias in the energy gain at the chip
location of Sgr~A.

Further analysis of the calibration lines indicates that the detection
efficiency in the low energy band ($\sim$ 2~keV) is almost flat across
each CCD, while it is significantly decreased in the high energy band
towards the top rows of each chip.  The effective area of the
telescope mirrors and the detection efficiency of ACIS were calculated
with the {\it mkarf} program in the Chandra Interactive Analysis of
Observations Software package (CIAO, Version 1.0) which, at the
present time, does not account for the positional dependence of the
detection efficiency due to the CTI effects.  We estimate that this
could cause the inferred temperatures to be systematically lower than
the true temperatures by $\sim$10\% at $kT_e=2$~keV. We do not remove
this bias in the analysis below.

To reject background events, we applied a grade filter to keep only
{\it ASCA} grades 0, 2, 3, 4, \& 6. We removed events from flaring
pixels using the ``flagflare'' routine written by T.~Miyaji.
Artificial stripes caused probably by hot pixels in the
frame-store region and by particles which hit on the CCD node
boundaries were also removed. Detailed procedures for cleaning
low-quality events are given in {\it
http://www.astro.psu.edu/xray/axaf/recipes/clean.html}.

\subsection{Image Flat Fielding}

Figure~\ref{figure:2}a shows a broad-band raw count image of the Sgr~A
region between 1.5 and 7~keV. To visualize better the complex
combination of extended and compact structures in the region, we apply
an adaptive kernel smoothing algorithm developed by Ebeling, White, \&
Rangarajan (2000) to the raw count image. This algorithm uses the
local density of events to determine the width of the Gaussian
smoothing kernel at each location across the image. However,
smoothing the raw count image convolves the diminished event density
due to gaps between the CCDs (which are already broadened by satellite
dithering) with real astrophysical structures.

To remove this instrumental effect, we create an exposure map and
adaptively smooth it using a map of the kernel widths used by the
algorithm to smooth the count map.  We then divide the smoothed image
by the smoothed exposure map to remove instrumental effects due to
mirror vignetting and the interchip gaps; this yields a flat-fielded
smoothed X-ray flux map.

As the effective area varies from region to region and from energy to
energy, this method is valid only for a narrow-band image.  In order
to make a broad-band image spanning 1.5--7~keV, we adaptively smooth
the 1.5--7~keV image to get a map of kernel widths, and use this
kernel map to smooth images and exposure maps in three narrow bands:
1.5--3.0~keV, 3.0--6.0~keV and 6.0--7.0~keV.  We then divide the
smoothed image in each band by the appropriate smoothed exposure map
to create three flat-fielded narrow-band flux maps (see
Figures.~\ref{figure:2}c-e).  Finally, we sum the narrow-band flux
maps to produce the broad-band map (see Figure~\ref{figure:2}b and
Figure~\ref{figure:3}). 
A typical Gaussian width for the smoothing is about 5 arcsec at the
center of Sgr A East.
Ratios among the three
bands give the hardness-ratio maps shown in Figure~\ref{figure:4}.

\subsection{Point Source Removal}

Several dozen point-like sources can be seen in
Figure~\ref{figure:2}. To study the diffuse components in and around
Sgr~A East, we removed these point-like sources.  Source detection is
based on a Mexican hat wavelet decomposition of the unsmoothed image
using the {\it wavdetect} program in the CIAO software package
\citep{Freeman00a}.  The source detection threshold was set at
$10^{-6}$, corresponding to $\sim 1$ spurious source per chip.  The
wavelet scales used were 1, $\sqrt{2}$, 2, $2\sqrt{2}$, 4,
$4\sqrt{2}$, 8, $8\sqrt{2}$, and 16 pixels. For the analysis of
extended features, we excluded all events lying within an 8$\sigma$
radius of each compact source, where $\sigma$ is the standard
deviation of the telescope point spread function at 1.5~keV at each
location. This should remove $> 90$\% of the point source photons.

\section{X-RAY PROPERTIES}

\subsection{Morphology}

Raw and smoothed broad-band (1.5--7~keV) X-ray images of the Sgr~A
radio complex are shown in Figures~\ref{figure:2}a--b.  Dozens of
point sources and complex extended structures are clearly seen, as
reported in Paper~I.  In Figure~\ref{figure:3}, we show the smoothed
broad-band X-ray image overlaid with radio contours from a 20~cm VLA
image of Sgr~A provided to us by F.\ Yusef-Zadeh.  The outer
oval-shaped contours are due to synchrotron emission from the
shell-like non-thermal radio source Sgr~A East, which may be a
supernova remnant \citep[SNR~000.0+00.0;][]{Ekers83,Green84}.  The
thermal radio source Sgr~A West, an HII region located in the central
parsecs of the Galaxy, appears on the western side of the Sgr~A
complex.  At 90~cm, the non-thermal emission from Sgr~A East is seen
to be absorbed by the ionized gas in Sgr~A West, which must therefore
be located along the line of sight between us and Sgr~A East
\citep{Yusef87,Pedlar89,Anantharamaiah99}.

Several bright compact X-ray sources can be seen in the vicinity of
Sgr~A West.  One of these sources is coincident to within $0\farcs35$
with the radio position of the compact non-thermal radio source
Sgr~A$^*$.  Analysis of the X-ray emission from this source and its
vicinity is reported in Paper~I.  In addition to the compact sources,
the Sgr~A West region shows bright diffuse X-ray emission superposed
on a broader region of extended emission which peaks $\sim1${\arcmin}
east of Sgr~A$^*$, and which appears to fill the central $\sim2$~pc of
the Sgr~A East radio shell (Fig.~\ref{figure:3}).  This broader
feature is especially conspicuous in the 6--7~keV band
(Fig.~\ref{figure:2}e), where the flux is dominated by iron-K line
emission (\S 3b).  Notably, no significant X-ray continuum or line
emission is seen from the location of the radio shell.  Based on its
spectral and spatial properties, we associate the source of this
diffuse X-ray emission with a hot optically thin thermal plasma
located within the Sgr~A East radio shell.

A curious linear feature $\sim0\farcm5$ long, which we refer to as the
`plume' in Paper I, can be seen extending (in projection) from the center of
Sgr~A East to the northwest. This feature is most obvious in the 3--6~keV
band (Fig.~\ref{figure:2}d).  The total count rate from the plume is only
5--10\% of that from Sgr~A East.  Although it is possible that the
plume is physically associated with Sgr~A East, our results on the spectral
properties of Sgr~A East given in this paper are unchanged
by the inclusion, or not, of the flux in the plume.
Looking in the 1.5--3~keV band (Fig.~\ref{figure:2}c), we clearly see
faint X-ray emission extending perpendicular to the Galactic plane in
both directions through the position of Sgr~A$^*$.  This emission is
not apparent in the 6--7~keV map, and therefore appears to be
unrelated to Sgr~A East.  The two-sided nature of this emission
suggests it might originate in some kind of bipolar outflow.  Results
of a detailed analysis of the X-ray emission from these structures
will be published elsewhere.

The total surface brightness in the direction of Sgr~A East is
$\sim4.3 \times 10^{-5}$ counts s$^{-1}$ arcsec$^{-2}$.  This was
measured by extracting counts from a circular region of radius
40\arcsec\ centered at (RA, Dec)$_{2000}$ $=$ ($17^h45^m44.1^s$,
$-29^\circ00\arcmin23\arcsec$), as shown in Figure~\ref{figure:2}a.
A total of 8,700 counts were extracted, corresponding to a count rate
of 0.22 counts s$^{-1}$.  The entire Sgr~A region is surrounded by
diffuse emission which varies considerably in intensity and spectrum.
It is thus difficult to determine exactly what background to subtract.
We chose to estimate the background using the counts within a
30\arcsec-radius circular region centered at ($17^h45^m34.0^s$,
$-29^\circ01\arcmin52\arcsec$).  A total of 708 counts were extracted in the 1.5--9~keV band
from the background region, yielding a surface brightness of $\sim6.1
\times 10^{-6}$ counts s$^{-1}$ arcsec$^{-2}$.  If this diffuse
emission is assumed to be roughly constant at Sgr~A East, then the
estimated background rate within the source region is
$3.1\times10^{-2}$ counts s$^{-1}$.  The net surface brightness and
count rate from Sgr~A East are then $\sim3.6  \times 10^{-5}$ counts
s$^{-1}$ arcsec$^{-2}$ and 0.18 counts s$^{-1}$.

From the smoothed images in different energy bands, we found that the
structure of the Sgr A East emission is spectrally dependent. The
half-power radius of the emission is $\sim 20\arcsec$ in the 6--7~keV
band compared to $\sim 30\arcsec$ in the lower energy bands.  The
concentration of hard emission towards the center can be seen in the
hardness-ratio maps of Figure 4, which show structure on scales of
10--20\arcsec\ towards the center of Sgr~A East.

\subsection{Spectra}

In order to quantitatively evaluate the spectrum of Sgr~A East, we
extracted source and background spectra from the regions described in
\S3.1.  As noted above, the diffuse emission surrounding the Sgr~A
region varies considerably in intensity and spectrum.  There is
therefore some uncertainty inherent in the background subtraction.
Using the spectrum extracted from a second background region centered
at ($17h45m50.6s$, $-29^\circ01'46''$), we found that
the net emission derived for the low energy band around 2~keV depends
somewhat on the choice of background, while the emission in the harder
bands is reasonably secure.  Our analysis indicates that none of the
scientific conclusions discussed in this paper are dependent on the
background subtraction. The background-subtracted count rate for the
40\arcsec-radius emission is 0.03, 0.12, and 0.03 counts s$^{-1}$ in
the 1.5--3.0, 3.0--6.0, and 6.0--9.0~keV bands, respectively.

The spectrum of Sgr~A~East, shown in Figure~\ref{figure:5}, exhibits a
continuum plus emission lines which give critical information on the
physical state of the plasma.  We first fit the spectrum with a thermal
bremsstrahlung model having four Gaussian emission lines of unspecified
energy, all absorbed by an interstellar medium having cosmic
abundances. The best-fit parameters are shown in Table~1.  The emission
lines can be attributed to the K$\alpha$ transitions of the helium-like
ions of sulfur, argon, calcium and iron.  The line width for each line
is consistent with being unresolved. The existence of the highly ionized
ions confirms the presence of an optically thin thermal plasma. The
equivalent widths of the lines are relatively small for the first three
elements ($EW \simeq 0.1-0.2$ keV) but very large for iron ($EW \simeq
3.1$ keV).  The continuum temperature is around 3 keV and the
line-of-sight column density is around $1 \times 10^{23}$ cm$^{-2}$,
equivalent to a visual absorption of $A_{\rm V} \simeq 60$~mag
\citep[we assume $N_{\rm H}$ = 1.79 $\times$
10$^{21}$ A$_{\rm V}${;}][]{Predehl95}.

The large equivalent width of the iron line suggests that the Sgr~A
East plasma is enriched in heavy elements. Since the high ionization
state of the iron K-line can be reproduced by a plasma in
collisional ionization equilibrium \cite{Masai94}, we fit the spectrum
to models of an isothermal plasma having variable elemental abundances
(MEKA; Mewe, Gronenschild \& van den Oord 1985), modified by
interstellar absorption. The model with solar elemental abundances
\citep{Anders89a} was rejected with $\chi^2 = 309$ (185 d.o.f.)\
because it does not reproduce the large equivalent width of the iron
line. The best-fit ($\chi^2 = 217$ with 184 d.o.f) was obtained using
heavy element abundances which are $\simeq 4$ times solar.  These
results are given in Table~2 and are shown as a solid line
in Figure 5.

In order to examine relative abundances among heavy elements, we allowed
only the iron abundance Z$_{\rm Fe}$ to be a free parameter, fixing
the hydrogen abundance at zero and holding other elements, Z$_{\rm
others}$ at their solar ratios. The best fit ($\chi^2$/d.o.f $=$
218/184) does not indicate that iron is more overabundant than other
metals : z$_{\rm Fe}$ $\equiv$ Z$_{\rm Fe}/$Z$_{\rm others}=$1.1(1.0--1.4).

Assuming energy equipartition between electrons and ions, the best-fit
model gives the following physical properties for the plasma, assuming
a spherical volume with radius 1.6 pc and an unknown filling factor
$\eta$: electron density $n_e \simeq 6 \eta^{-\frac{1}{2}}$ cm$^{-3}$,
gas mass $M_{\rm g} \simeq 2 \eta^{\frac{1}{2}}$ M$_{\odot}$, X-ray
luminosity $L_{\rm x} \simeq 8 \times 10^{34}$ erg s$^{-1}$ in the
2--10 keV band\footnote{Here and elsewhere, $L_{\rm x}$ is the
absorption-corrected luminosity in the 2--10 keV band unless otherwise
noted.}, and total thermal energy $2\times10^{49}
\eta^{\frac{1}{2}}$ ergs.

For a quantitative analysis of the radial dependence of the spectra,
we divided the Sgr~A East region into two concentric annuli, each with
a $20''$ width (Figure~\ref{figure:2}a), and obtained X-ray spectra
from each annulus (Figure~\ref{figure:6}). We fitted the spectra with
the absorbed thermal bremsstrahlung plus Gaussians model fixing the
line energies to be those given in Table~1. The best-fit
parameters are plotted in Figure~\ref{figure:7}. The most striking
spatial variation is that the equivalent width of iron in the inner
region is elevated by a factor of 1.5 compared to the surrounding
region, while the equivalent widths of sulphur, argon and calcium are
consistent with being uniform. The electron temperature also appears
constant between the two regions. The relative abundance z$_{\rm Fe}$
of iron compared to other metals in the MEKA model assuming the
temperature and the absorption are the same in the two regions shows
2.2(1.7--2.8) and 0.8(0.7--1.0) for the inner and outer regions,
respectively ($\chi^2$/d.o.f $=$ 331/295). Even if we allow
temperature and absorption to vary, the abundance gradient is present
with $3\sigma$ confidence. We thus conclude that iron is concentrated
in the interior region by a factor of $\sim2$, while the lighter
elements are spatially roughly homogeneous.

\section{DISCUSSION}

\subsection{Origin of Sgr~A East}

The X-ray spectrum enriched by heavy elements suggests that the X-ray
plasma is dominated by supernova ejecta. The small gas mass of $2
\eta^{\frac{1}{2}}$ M$_{\odot}$ and thermal energy $\sim10^{49}$~ergs
are consistent with the ejecta by a single SN explosion. These results
straightforwardly supports the long-standing hypothesis that Sgr
A~East is a single SNR \cite{Ekers83}.

Rho and Petre (1998) defined a new class of composite SNRs showing
centrally concentrated thermal X-rays lying within a shell-like
non-thermal radio structure. They called these objects ``mixed
morphology'' (MM) supernova remnants and identified 19 members of the
class.  Bamba et al. (2000) reported that G~359.1$-$0.5, which, like
Sgr A~East, is in the Galactic Center region, shares the defining
features of MM SNRs.  With the centrally concentrated X-ray emission
we find here, and its well-established non-thermal radio shell,
Sgr A~East becomes a new member of the class of MM SNRs.  Additional
evidence for this conclusion is provided by the 1720-MHz OH masers
present in the Sgr A East shell \citep{Yusef96}.  Such masers are
often detected in MM~SNRs \citep{Green97}.

Figure~\ref{figure:8} plots a histogram of MM SNR sizes including
Sgr~A~East. No historical SNR, such as Tycho, is a MM-SNR, suggesting
that Sgr~A East is not very young with an age roughly $t>10^3$ yrs
old. No MM~SNRs are seen in the LMC or SMC, in which interstellar
matter is known to be less dense than that of the
Galaxy. Figure~\ref{figure:8} shows that Sgr~A East is the smallest of
the MM SNRs, probably indicating Sgr~A~East is evolving into ambient
materials denser than those for usual MM SNRs.  One MM SNR, W49~B
(G~43.3-0.2), has a size similar to that of Sgr~A East. The basic
properties of Sgr~A East and W49~B are summarized in Table~3. The X-ray properties of W49~B and Sgr~A East are also
similar \citep{Pye84,Smith85,Fujimoto95,Sun00,Hwang00}: (1) the
equivalent width of the iron K-line is very large indicating a plasma
enriched in heavy elements; (2) narrow-band X-ray imaging indicates a
spatial gradient of elemental abundance where the distribution of iron
is more compact than those of the lighter elements; and (3) X-ray
spectra indicate an X-ray plasma in collisional ionization
equilibrium.  The close resemblance between W49~B and Sgr~A East
confirms the notion that Sgr~A East is best interpreted as a SNR and
is not a unique object. 

Exotic hypotheses invoking the nearby MBH, such as an explosion
inside a molecular cloud, possibly due to a tidally induced
catastrophic event occurring $\sim10^{5}$ yrs ago within 10
Schwarzschild radius of the MBH \citep{Yusef87,Khokhlov96}, are
not likely to reproduce the metal-rich plasma because the explosion
is driven by gravity rather than by a nuclear reaction.  The other
exotic hypothesis -- near-simultaneous explosions of $\sim40$~SN
\citep{Mezger89} -- can produce heavy elements, but this hypothesis
is very strongly constrained by the short expansion time of the shell,
and by the absence of a massive, young stellar cluster near the center
of the shell.  In summary, the most straightforward hypothesis for the
origin of Sgr~A East is a single supernova.  Alternative hypotheses are
hard pressed to quantitatively reproduce both the X-ray and radio
properties.

\subsection{Age}

Mezger et~al. (1989) estimated the age of Sgr~A East to be
$t\sim$7,500 yrs assuming that a SNR evolves into a bubble in a giant
molecular cloud formed by a stellar wind from the progenitor. Uchida
et~al. (1998) independently derived an age of a few $10^4$ yrs based
on the shear effect due to the differential Galactic rotation. These ages
are roughly close to that of W49~B \citep[$10^{3-4}$ yrs: see][and
reference therein]{Moffett94} and is consistent with that of a typical
MM SNR ($t>10^3$ yrs). These arguments support an age of Sgr~A East of
approximately $10,000$ yrs. This age is also consistent with the
observed X-ray properties as discussed in the later sections.

Note that the age is too short to form the compact HII regions which
are probably $10^{5-6}$ yrs old seen in radio continuum maps
(Figure~\ref{figure:3}), We thus support the argument by Serabyn et
al. (1992) that Sgr~A East did not stimulate star formation running
along the chain of the compact HII regions. 

\subsection{Plasma Diagnostics}

Since the electron density of the X-ray plasma is $\sim 6
\eta^{-\frac{1}{2}}$ cm$^{-3}$, the ionization parameter ($n_e t$
$\simeq$ $2\times10^{12}$ $\eta^{\frac{1}{2}}$ cm$^{-3}$~s assuming
$t=1\times10^{4}~yr$) is near the characteristic timescale for
realizing collisional equilibrium \citep{Masai94}. For this density
and temperature, the electrons and ions should reach energy
equipartition in $10^3$ yrs \citep{Spitzer62}, which is an order of
magnitude shorter than the age estimated for Sgr~A East. The sound
crossing length, an effective length for heat conduction, is
$\sim8$~pc at $10^4$~yrs, which is longer than the radius of the
plasma ($\sim1.6$~pc). The radiative cooling time of the plasma is
$10^{6}$ yr. Our phenomenological success in fitting an isothermal
MEKA model (a single temperature plasma in collisional ionization
equilibrium) to the spectrum (\S 3.2, Figures 5 \& 7) is completely
consistent with those plasma conditions, which now have a physical
foundation. Therefore, the total energy of the plasma derived by
assuming energy equipartition, $10^{49}$~ergs, is likely to be
correct.

\subsection{Stellar Progenitor}

Recall from \S 3.2 that the best-fit relative abundance gives $z_{\rm
Fe} = 1.1$ and that, averaged over the remnant, iron is as abundant as
the other heavy elements. Type Ia explosions give the highest ejection
of Fe/Ni-group elements (z$_{\rm Fe}\sim$ a few) while Type II
explosions on average give lower abundances \citep[$z_{\rm Fe} \simeq
0.5${;} e.g.{,}][]{Tsujimoto95a}. Among Type II explosions, the iron
abundance is about unity for a progenitor mass of
$M=13$--20~M$_{\odot}$, but is lower by an order of magnitude for
$M=40$--70 M$_{\odot}$ \citep[e.g.\,][]{Nomoto97}. Therefore, the
observed iron abundance corresponds to that predicted for a Type~II
explosion in a progenitor star of mass 13--20~$M_{\odot}$.  A supernova
from a progenitor in this mass range ejects $1-5$ M$_\odot$ of
material, which is consistent with the $2$~M$_\odot$ of plasma deduced
in \S 3.2 assuming a high volume filling factor ($\eta \sim 1$).
The explosion energy of this type of SN is $10^{51}$ ergs, which is one
or two orders of magnitude smaller than the total energy suggested by
Mezger et al. (1989).

A Type II explosion is believed to produce a neutron star which is
often kicked up to a fairly high velocity \citep[For a fast case,
$\sim700$~km~s$^{-1}$, ][]{Cordes98}.  The kicked neutron star might
thus have traveled a distance of 700~km~s$^{-1}$ $\times$ $10^4$~yr
$=$ 7~pc.  Since the ambient material is dense (\S 4.5), the neutron
star running through it could produce a strong bow shock, which might
account for the observed linear X-ray feature, the "plume".  The distance
traveled is, in projection, about 3~arcmin $\times$ sin~$\theta$,
where $\theta$ is the angle between the neutron star's motion and the
line of sight.  The observed linear feature, with a length of
$\sim$0.5 arcmin, could be consistent with a bow shock tail of the
neutron star if its velocity vector is close to the line of sight.  A
future {\it Chandra} observation with a very long exposure would
probably be the best way to determine whether a neutron star is
present at the tip of the plume.

\subsection{Dynamical Evolution I: Forward Shock and Ambient 
Environment}

Two scenarios have been put forth to explain the formation of MM SNRs:
cloud evaporation or `fossil radiation'.  In the evaporation scenario,
the enhanced interior X-ray emission arises from gas evaporated from
cold interstellar clouds that were enveloped by the SNR
\citep{White91}.  Clouds too small to upset the overall forward-shock
propagation and of sufficient density to survive passage through the
shock provide a reservoir of material inside the remnant cavity.
Their subsequent evaporation increases the density of the hot interior
and hence the X-ray emissivity. The clouds must be numerous and have a
small filling factor in order to produce the X-ray emission without
affecting the shock dynamics. In the fossil radiation scenario, the
SNR moves into a dense and less clumpy ISM which is almost
completely snowplowed into a dense shell by the advancing shock
\citep{McKee75}.  The SNR has a luminous plasma shell associated with
the forward shock but this shell has cooled to low temperatures which
would be undetectable in X-rays through the line-of-sight absorbing
material towards the Galactic Center.  The presence of this shell is
instead revealed by radio emission.  The ejecta heated by the reverse
shock is thus detected as `fossil' thermal radiation within an
invisible shell \citep{McKee74}. Here, the X-ray emitting plasma is
dominated by the ejecta, whereas it is dominated by ordinary
interstellar material in the cloud formation scenario.  The highly
metal-enriched spectrum found in Sgr A~East favors the fossil
radiation scenario, i.e., ejection into a dense and homogeneous
ambient environment.

Metzer et al. (1989) and Uchida et al. (1998) have already discussed
the dynamical evolution of the Sgr~A East SNR on the basis of its radio
properties.  The picture of the ambient material given by Metzer
et~al. (1989) is that of a huge wind bubble formed in a giant molecular
cloud, while Uchida et al. (1998) consider a less dense and homogeneous
gas.  The X-ray properties then support the picture of Uchida
et~al. (1998).   Sgr~A East is likely to be in the region of non-solid-body
rotation near the gravitational center, Sgr~A* \citep[e.g.,][]{Uchida98},
where the rotation time-scale is $\sim 4\times10^{4}$~$(R/1~$pc$)^{1.5}$~yrs
at Galactrocentric radius $R$ \citep[see rotation curve in][]{Lugten86}.
Differential rotation tends to shear the ISM and to smooth it out in
a few rotation cycles ($\sim 10^{5}$~yr). These considerations favor the
homogeneous ambient environment.

The non-thermal radio emission from Sgr~A East in the direction of
Sgr~A West is very faint, mainly due to the heavy absorption by the
spiral-shaped group of thermal gas streams called ``the mini- spiral''
\citep{Yusef87,Pedlar89,Anantharamaiah99}. These authors also reported
that a turn-over (absorption) occurs at 90~cm from both Sgr~A East and
West, which suggests the presence of an ionized gas halo extending over
the Sgr~A complex.  This ionized halo has an implied emission measure
of $\sim10^5$~pc cm$^{-6}$ and an optical depth at 90~cm of around
unity.  Anantharamaiah et~al. further suggest that it has an angular
extent of $\sim4'$($\sim$~9~pc), an electron density of $10^{2-3}$
cm$^{-3}$, and an electron temperature of maybe 0.5--1 eV.  The radial
extent of the ionized gas halo nearly corresponds to that of the
non-solid-body rotating region ($\sim10$~pc). Therefore, the most
straightforward idea is that an ionized gas halo is filling the
non-solid-body rotation region with a nearly homogeneous density, and
that Sgr~A East is expanding into this ionized gas halo
(Figure~\ref{figure:9}).

Another potential candidate for the ambient material is the ``diffuse
X-ray plasma'', which extends across the inner few hundred parsecs of
the Galactic Center \citep{Koyama89}.  The temperature of the diffuse
X-ray plasma is as high as 10~keV, so the Mach number is probably less
than unity and the SNR will expand without interaction.  This is
likely to be inconsistent with the existence of the forward shock
evidenced by the bright radio shell (Figure~\ref{figure:3}).  This
argument is also consistent with the alternative interpretation for
the diffuse X-ray emission: that the emission does not originate from
a thin thermal plasma but is due to charge-exchange interactions of
low-energy cosmic-ray heavy ions \citep{Tanaka00}.

The dynamical evolution of Sgr~A East in the Galactic latitudinal
direction (perpendicular to the plane) might not be affected by
Galactic rotation or by the strong and maybe vertical magnetic fields
\citep[][and references therein]{Morris96}, so we can directly compare
the latitudinal extent of Sgr~A East to that predicted by the simple
theory for dynamical evolution of a SNR \citep{Lozinskaya92}. The
theory predicts that, at an age of $10^4$ yr, Sgr~A East is presently
in its radiative phase.  Although the dynamical evolution in the
radiative phase of the theory is somewhat uncertain, the radius is
predicted to be 6 $(t/10^4 {\rm yr})^{0.31}$ $(n_{\rm e}/10^3 {\rm
cm^{-3}})^{0.25}$ pc, where the total mass, the initial velocity of
the ejecta, and elemental abundances are assumed to be 2~M$_{\odot}$,
$10^4$ km s$^{-1}$, and solar, respectively. If the electron density
$n_{\rm e}$ of the ambient gas is $10^{3}$~cm$^{- 3}$, corresponding
to the denser estimate for the ionized gas halo, the radius roughly
agrees with that of the radio shell ($\sim$2.9~pc; see
Figure~\ref{figure:3}). The large column density of the ionized gas
halo, $3\times10^{22}$ H cm$^{-2}$ (R/10 pc) (n$_e$/$10^3$ cm$^{-3}$)
is presumably related to the observed discrepancy between the optical
extinction to the GC ($A_{\rm V}$ $\simeq$ 30~mag or $N_{\rm H}$
$5\times10^{22}$ cm$^{-2}$) and the X-ray absorption column density
($N_{\rm H}$ $\simeq$ $10\times10^{22}$ H cm$^{-2}$), as discussed in
Paper~I.  The discrepancy can be explained if the ratio of dust grains
to atoms is much smaller in the ionized gas halo than in the
foreground gas: infrared light is extincted mainly by the foreground
gas while X-rays are absorbed by both the foreground and the
ionized halo gas around Sgr A East.

Lozinskaya predicts the shock velocity for the blast wave to be as
slow as $2\times10^2$~$(t/10^4 {\rm yr})^{-0.69}$ km s$^{-1}$, and
that the post-shock temperature is as warm as $\sim70$~eV.
Ultraviolet and soft X-rays from the warm plasma should be completely
absorbed, which is consistent with the non-detection of X-rays from
the radio shell. The cooling time of the warm plasma is shorter than
one year \citep{Raymond77}, which is at least four orders of magnitude
shorter than the SNR age. The rapid cooling implies a cooled thermal
shell in the forward shock region. In fact, the cooled thermal shell
was detected as a dust ridge surrounding the non-thermal radio shell
by Mezger et~al. (1989) (Figure~\ref{figure:1}).

Mezger et~al. also found that the dust ridge in the eastern half is so
dense that molecular line emission is detectable, thus it is called
the curved molecular ridge (hereafter ``the molecular ridge'').  They
found that the total gas mass of the dust/molecular ridge was
$6\times10^4$ M$_{\odot}$.  With the assumption of a high filling factor,
$\sim1$, the total gas mass of the molecular ridge was independently
derived as $1.5\times10^5$ M$_{\odot}$ using the molecular-line
observations \citep[][and reference therein]{Coil00}.  This is 20-50
times larger than the mass of $3\times10^3$~$(r_{\rm
SNR}/2.9~$pc$)^{3}$ $M_{\odot}$ that the Sgr~A East shell is able to
accumulate from the ambient matter having $n_e=10^3$ cm$^{-3}$.

The mass discrepancy reminds us of the arguments by Mezger
et~al. \citep[1989, see also][and reference therein]{Yusef00}, that
the eastern part of Sgr~A East has been expanding into the
$+$50~km~s$^{-1}$ cloud, sweeping up gas at the western edge of the
cloud, compressing it and forming a substantially denser ridge at
the eastern edge of Sgr~A East than on the western side (see
Figure~\ref{figure:1}).  In this case, the compressed gas mass is
mostly included in the western shell, the mass of which becomes
$\sim10^5 M_{\odot}$ because the density of the $+$50~km~s$^{-1}$ cloud
($10^5$~cm$^{-3}$) is about a hundred times higher than that of the
ionized gas halo \citep{Coil00}.  The mass of $10^5$ M$_{\odot}$ is
consistent with the mass presently surrounding the shell. The cloud
should also brake the eastern shell of Sgr~A East, the speed of which
is reduced by a factor of 10 to ($\sim20$~km~s$^{-1}$).  The small
velocity dispersion of the molecular ridge \citep[a few $\times$ $10$
km~s$^{-1}$][]{Coil00} is highly consistent with this picture.

As the eastern side of the Sgr~A East shell should be located further
away from the gravitational center at Sgr~A* than the western side, and
the Galactic rotation timescale is proportional to $\sim1/R^{1.5}$,
the large $+$50~km~s$^{-1}$ cloud can probably persist for as long as
$10^6$~yr before being sheared so much that it becomes a continuous
stream of gas rather than a well-defined cloud.  We therefore propose
a picture in which most of the Sgr~A East shell is expanding into the
ionized gas halo where the density is $\sim~10^3$~cm$^{-3}$, but the
eastern side has encountered the much denser $+$50 km s$^{-1}$ molecular
cloud.  The implication of this model is that the expansion has proceeded
further toward the west than toward the east.

While the thermalized component associated with the shock front
experiences large radiative cooling by atomic processes, the
non-thermal component observed as a non-thermal radio shell should
lose a large amount of energy by the synchrotron process operating in
the strong milliGauss magnetic field. The non-detection of X-rays from
the radio shell and the steep synchrotron radio spectrum are both
understood simply in terms of radiative energy loss \cite{Pedlar89}.
Melia et al. (1998) suggested the possibility that infrared light from
Sgr~A West might be Comptonized within the radio shell to the hard X-ray
band with a fairly high luminosity of $L_{\rm x}$ $\simeq$
$2\times10^{35}$ ergs s$^{-1}$. This is at least two orders of
magnitude higher than the observed X-ray emission from the radio
shell, so this model is apparently not applicable.

\subsection{Dynamical Evolution II: Reverse Shock and SN Ejecta}

 The interaction of the forward shock with the dense surrounding
medium will have produced a reverse shock propagating back into the
expanding ejecta.  Since the electron temperature of the X-ray plasma
is $\sim$2~keV, we expect the shock velocity $v_s$ to be $\sim1,000$
km s$^{-1}$, which is about five times faster than the present shock
velocity expected for the forward shock ($\sim$200~km~s$^{-1}$, see \S
4.5).  This discrepancy indicates that the reverse shock has already
propagated back through the bulk of the X-ray emitting plasma, and the
high velocity suggests that the reverse shock has already arrived at
the center in an early phase and that the ejecta have been
thermalized.

 In SN explosions predicted by standard nucleosythesis theories, the
lighter elements like S, Ar, and Ca are expected to be ejected faster
than iron \cite[e.q.{,}][]{Nomoto97}.  If mixing is not effective, the
stratification of the elements may show up in the ejecta plasma.  In
Figure~\ref{figure:4}, we saw that the spatial extent in the 6--7 keV
band is smaller than those in the lower energy bands. The spectrum in
Figure \ref{figure:5} shows that the flux from the iron K-lines
dominates the 6--7~keV band, while the sulphur and calcium lines
enhance the 1.5--3 and 3--6 keV bands, respectively. The energy
dependence of the spatial extent suggests that the heaviest elements
are more compactly distributed than the lighter ones.  The spatial
distribution in every energy band shows a centrally concentrated
morphology (Figure~\ref{figure:2}), which indicates that the elements
co-exist at the inner region. The co-existence can be simply explained
if the elements in the supernova ejecta are not perfectly stratified
and well mixed in the inner region.

 The velocity dispersion of the 2~keV plasma is $\sim$620
$\sqrt{1/A}$~km~s$^{-1}$, where A is the atomic mass. So the mixing
length is estimated to be $\sim6.3$ $\sqrt{1/A}$ $(t/10^4\rm yrs)$ pc,
with which we can estimate the mixing angular scale on the sky of
$^{32}$S and $^{56}$Fe to be $29\arcsec$ and $22\arcsec$,
respectively. These values are very similar to the observed scale
radii of $\sim30\arcsec$ and $\sim20\arcsec$ in the 1.5--3 and 6--7
keV bands. Hence the estimated age of 10,000~yrs is long enough for
the light materials, like S, to reach the center but is too short for
iron to reach the outward light-elements layer. The observed spatial
gradient of elements could be consistent with the elemental
stratification predicted by standard nucleosythesis theories if
significant mixing occurs after the ejecta were thermalized.

 The observed brightness of the X-ray emission is centrally
concentrated, indicating that the outer region of the ejecta plasma is
sparse. In fact, as the forward shock compresses the ambient
materials, a hot cavity is expected to be formed inside of the forward
shock layer \cite{Chevalier82}.
The ejecta plasma seen in X-rays might currently be undergoing mixing
as well as diffusing into the hot cavity.

\subsection{The Ionized Gas Halo and Its Irradiating Source}

The radio non-thermal shell still exists, indicating that the forward
shock velocity of the shell is faster than the sound velocity of the
ambient material, 'the ionized gas halo'. Therefore, the temperature
of the ionized gas halo is substantially lower than $\sim$70~eV. If
the halo is collisionally ionized, the temperature should be higher
than $\sim10$~eV. The radiative efficiency is known to have a peak
around 20~eV. So, in order to keep the temperature in the range
between 10 and 70 eV, kinematic heating of $10^{43}$
($n_e$/$10^3$~cm$^{-3}$) ($R$/10~pc)$^3$ ergs s$^{-1}$ is
required. Such an energetic source can not be found in the Galactic
Center region, so the gas should have rapidly cooled down in 1~yr
which argues strongly against collisional ionization heating.

An alternative is that photo-ionization ionizes the gas halo, in which
case the kinematic temperature is much lower than $10$~eV. Since the
column density is $3\times10^{22}$ H cm$^{-2}$ ($R$/10 pc)
($n_e$/$10^3$ cm$^{-3}$), ultra-violet photons can contribute the
ionization of only a local and small portion of the ionized gas
halo. X-rays above $\sim$1~keV are expected to be the main contributor
to photo-ionization. Although the photo-ionization cross section is
highly dependent on the spectral slope of the irradiating X-ray
source, $L_{\rm X}\simeq10^{40}$($n_e/10^{3}$~cm$^{-3}$)($R$/10~pc) is
required to ionize most of the hydrogen atoms in the non-solid-body
rotation region \cite[see Figure~3 in][]{Kallman82}. However, no
persistent X-ray source brighter than of an order $10^{36}$ ergs
s$^{-1}$ is located in the Galactic Center region
\citep[e.g.,][]{Pavlinsky94}.

The absence of a predicted bright X-ray irradiator reminds us of the
study of Sgr~B2 a unique X-ray reflection nebula near the Galactic
Center \cite{Koyama96a,Murakami00}. They found fluorescent X-ray
emission from cold iron atoms in molecular clouds, possibly due to
irradiation by X-rays from Sgr~A*, which was bright $10^3$ yrs ago,
but is presently dim.  By applying a detailed model of an X-ray
reflection nebula to the Sgr~B2 molecular cloud, they estimated the
X-ray luminosity of Sgr~A* 300 yrs ago to be $\sim3\times10^{39}$ ergs
s$^{-1}$, which is comparable with that required to ionize the gas
halo ($\sim10^{40}$ ergs s$^{-1}$). The recombination time in a $10^3$
cm$^{-3}$ gas is as short as $\sim300$ years. Thus the high-ionization
fraction in the ISM rotating in the non-solid-body rotation region
could be due to the past activity of Sgr~A* at about $10^{2-3}$ years
ago and the ISM is presently in a recombination-dominated phase. Thus,
the existence of the ionized gas halo surrounding Sgr~A* supports the
scenario that Sgr~A* experienced AGN activity in the recent past
\citep{Sunyaev93,Koyama96a,Murakami00,Murakami01}. Note that the gas
halo would have been neutral before the recent AGN activity if no
other irradiator was present.

\subsection{Comments on A Possible Relation Between Sgr A~East and Sgr~A*} 

The non-thermal radio emission from Sgr~A East in the direction of
Sgr~A West is heavily absorbed by Sgr~A West
\citep{Yusef87,Pedlar89}. This fact convincingly indicates that, along
the line-of-sight, Sgr~A~West lies in front of the Sgr A~East shell.
However, the distance between Sgr A~West and Sgr A~East along the
line-of-sight is uncertain. It is quite possible that they lie at
nearly identical distances, in which case the front edge of the
expanding Sgr~A East shell has probably reached and passed through
Sgr~A West \citep[][and reference therein]{Morris96}. This
configuration is simply illustrated in Figure 9, see Yusef-Zadeh
et~al. (2000) for a more thorough discussion.

 Four observations support this hypothesis. First, if a structure
of size much smaller than the scale of Sgr~A East is being
overrun by the forward shock of the SNR, a bow shock should form
in the direction of explosion of Sgr~A East \citep{McKee75}.
Such features may have been found from the molecular ring in
recent radio studies \citep{Yusef00}.  Second, if Sgr~A West is
embedded within the non-thermal shell of Sgr~A East, the radio
emission from the front side of the non-thermal shell could escape
being absorbed by the thermal ionized gas in Sgr~A West.
Yusef-Zadeh et~al. (2000) report that there is indeed faint,
non-thermal emission present toward the thermal ionized gas at
90~cm.  Third, if Sgr~A East is located very near Sgr~A*, it
should be sheared by the non-solid Galactic rotation \citep{Uchida98}.
Yusef-Zadeh et~al. (1999) reported that the kinematics of the OH
maser spots associated with Sgr~A East are consistent with such
shear motions and with a separation less than 5 pc.  Fourth, we
found that, except for the $+$50 km s$^{-1}$ molecular cloud, the
ambient matter surrounding Sgr A East was probably homogeneous,
giving rise  to the dust ridge that appears all around the
shell.  However, the dust ridge surrounding most of Sgr A~East
(Figure 1) is likely to be missing in the vicinity of Sgr A~West
\citep{Mezger89, Dent93}.  This discrepancy suggests that the dust
ridge has overrun Sgr A~West and has been dynamically disrupted there
by merging with the circumnuclear disk or, to a lesser extent, by
accretion onto the MBH at Sgr~A*.  It is thus likely that Sgr~A*
lies within `the hot cavity' inside the matter compressed by the
shock front.  The radio image in projection is consistent with
this scenario (Figure~\ref{figure:1}).  However, it is less obvious
that Sgr~A* is in direct contact with the hot medium within the shell
because of competition from the winds emanating from the massive
emission-line stars in the central cluster ({\it c.f.}, Paper I).

 Assuming that the dynamical force of stellar winds in the central
parsec is negligible, then, using Bondi-Hoyle theory \cite{Bondi44},
we can roughly estimate the expected  accretion rate of gas onto the
MBH as the dust ridge compressed by the SNR shock passes Sgr~A*.
When the dust ridge passes by, the radius within which material falls
onto the MBH with a mass of $M_{\rm MBH}$ is regulated principally
by the velocity of the dust ridge, $v_{\rm dust}$, according to
\begin{equation}
R_{\rm Bondi} = 0.5~{\rm pc}~(\frac{v_{\rm dust}}{200~{\rm km}~{\rm s}^{-1}})^{-2} (\frac{M_{\rm MBH}}{2.6\times10^{6}~{\rm M}_{\odot}})
\end{equation}
with a corresponding accretion rate
\begin{equation}
\dot{M}_{\rm bondi} = 0.02~{\rm M}_{\odot}~{\rm yr}^{-1} (\frac{n_{\rm
dust}}{{\rm 4,000}~{\rm cm}^{-3}}) (\frac{v_{\rm dust}}{200~{\rm km}~{\rm s}^{-1}})^{-3}
(\frac{M_{\rm MBH}}{2.6\times10^{6}~{\rm M}_{\odot}})^2,
\end{equation}
where $n_{\rm dust}$ is the mean electron density assuming that the
ambient material of $10^3$ cm$^{-3}$ is compressed by a factor of
four, and $v_{\rm dust}$ is the expanding velocity of the dust ridge
assuming the same as the shock velocity expected for the forward
shock. The crossing time of the dust shell is
\begin{equation}
t_{\rm crossing} \simeq 1\times10^3~{\rm yrs} (\frac{\Delta R}{0.24~{\rm pc}}) (\frac{v_{\rm dust}}{200~{\rm km}~{\rm s}^{-1}})^{-1}
\end{equation}
where $\Delta R$ is the width of the dust ridge assuming one
twelfth of the radius of the ridge. The total accreted mass is roughly
estimated to be $20~{\rm M}_{\odot}~(R_{\rm bondi}/0.5~{\rm
pc})^{2}$. The predicted mass accretion of around $10^{-2}~{\rm
M}_{\odot}~{\rm yr}^{-1}$ is comparable to the Eddington limit of
$5\times10^{-3}~{\rm M}_{\odot}~{\rm yr}^{-1}$, suggesting Sgr~A*
could have been as bright as the Eddington luminosity of
$3.4\times10^{44}$ ergs s$^{-1}$ for $\sim 1\times10^3~{\rm yrs}$.
Hence, the dust ridge, the ISM compressed by the forward shock of
Sgr~A East, might be dense enough to activate the MBH to Seyfert or
quasar luminosity level in the recent past.

It is possible, however, that stellar winds might  impede the accretion of
the dust ridge.  The winds from the central cluster of early-type
emission-line stars have an estimated integrated mass loss rate
$\dot{M}_{\rm winds} \simeq 4\times10^{-3}$ M$_{\odot}$~yr$^{-1}$ with
an average wind velocity $v_{\rm winds} \simeq$ 700~km~s$^{-1}$,
inferred from hydrogen Br$\gamma$ line observations \cite{Yusef00}.
At a distance of half a parsec, roughly the size of the stellar
cluster, the ram pressure from the winds
\begin{equation}
\rho_{\rm winds} v_{\rm winds}^2 \simeq 1\times10^{-6}~{\rm dynes}~{\rm cm}^{-2}
\end{equation}
is comparable to the estimated ram pressure of the dust ridge,
\begin{equation}
\rho_{\rm dust} v_{\rm dust}^2 \simeq 3 \times 10^{-6}~{\rm dynes~cm}^{-2}(
\frac{n_{\rm dust}}{4\times10^3~{\rm cm}^{-3}})
(\frac{v_{\rm dust}}{200~{\rm km~s}^{-1}})^2.
\end{equation}
The stellar winds could thus impede the captured dust ridge from
accreting onto the MBH. The accretion rate would then be reduced by an
unknown factor.

We suggest that the dense dust ridge arrived near the MBH and its
surrounding central cluster about 1--2$\times10^3$ yrs ago, producing
an increased external ram pressure that overwhelmed the pressure of
the stellar winds which led to a large jump in the accretion rate,
triggering the AGN activity that ionized the gas halo around Sgr~A
East and is illuminating the more distant molecular cloud Sgr~B2.  The
ridge passed by the MBH one or two hundreds yrs ago and the luminous X-rays
associated with the AGN activity faded away. The accretion from the
dust ridge was completely finished before the beginning of X-ray
astronomy and the MBH at Sgr~A* has never been detected in X-rays
before the {\it Chandra} era \citep[Paper~I,][for a history of X-ray
observations]{Maeda96}. Sgr~A* is then currently embedded in the hot
cavity of the Sgr~A East SNR, in which little accretion is possible and
the MBH is observed in a very quiescent state with $L_{\rm x} \simeq
10^{33}$ ergs~s$^{-1}$ (Paper~I).

\section{CONCLUSION \& SUMMARY}

Using the ACIS instrument on board {\it Chandra}, we have spatially
resolved for the first time the X-ray emission of the shell-like
non-thermal radio source Sgr~A East (SNR 000.0$+$00.0) from the other
complex X-ray structures present in the Galactic Center. We find the
X-ray emission is concentrated within the central $\simeq 2$ pc radius
of the larger $\sim 6-9$ pc radio shell. The spectrum clearly
originates within an optically thin thermal plasma with strong
K$\alpha$-lines from highly ionized heavy atoms indicating an
overabundance of heavy elements several fold above solar levels.  The
elemental abundances show a spatial gradient: the distribution of iron
is more compact than the lighter elements.  The morphology (Mixed
Morphology), energetics
($\sim2\times10^{49}$~$\eta^{\frac{1}{2}}$~ergs~s$^{-1}$), and mass
($2 \eta^{-1/2}$ M$_\odot$) are all consistent with a single supernova
remnant origin. The relative abundances between heavy elements favors
Sgr~A East originating from a Type~II SN of a 13--20 M$_\odot$
progenitor star. An exotic origin related to the Sgr~A$^*$ massive
black hole is not required.

 While detailed modeling of the structure is subject to considerable
uncertainties, we have evaluated a simplified model for the dynamical
evolution of Sgr~A East as a SNR formed about 10,000 years ago by a
Type~II supernova. In this model, the ejecta expanded into a
homogeneous and dense interstellar medium of $10^3$~cm$^{-3}$ that
pervades the central $\sim10$~pc around the Galactic Center. We see
the SNR today when the forward shock is relatively slow, explaining
the absence of hard X-rays associated with the larger radio shell. The
reverse shock was relatively fast, forming a hot plasma in the
interior of the remnant that is still visible in the hard X-ray
band. The result is a rare subtype of SNR: a very compact ``mixed
morphology'' remnant. Of all known SNRs, only W49~B appears to be
similar to Sgr~A East.

 Sgr~A East is interacting with the $+50$~km~s$^{-1}$ molecular cloud
on its eastern side.  Its relationship to Sgr~A West is still
controversial. We suggest that the dust ridge compressed by the
forward shock reached Sgr~A$^*$ $\sim10^3$ yrs ago. The passage of
the dust ridge may have led to increased accretion onto the MBH and
triggered nuclear activity, the remains of which are observed today as
the ionized gas halo surrounding Sgr~A$^*$ with a radial extent of
$\sim9~pc$ and the X-ray reflection nebula, seen in the more distant
Sgr~B2 molecular cloud. The MBH at Sgr~A$^*$ currently might lie
within the hot cavity of the Sgr~A East SNR, and is thus starved of
accreting material, explaining the extremely low X-ray luminosity from
Sgr~A$^*$ reported in Paper~I. Thus, the Galactic Center might be a
laboratory in which a single supernova remnant has controlled the
activity of the nuclear MBH.  This could be a realization of the broad
concepts relating nuclear starburst to MBH accretion activity in
galactic nuclei \citep[e.g.,][]{Heckman00}.

\acknowledgments 

This observation was performed as part of the guaranteed time
observation (GTO) program awarded to the ACIS development team led by
Gordon Garmire. We express our thanks to the entire {\it Chandra} team
for their many efforts in fabricating, launching, and operating the
satellite, and for their work in developing software for calibrating
and analyzing the data. Rashid Sunyaev shared valuable thoughts with
us on Galactic Center issues. Steven Reynolds, Jack Hughes, and other
attendees of the 11th Maryland conference `Young Supernova Remnants'
honoring the retirement of Steve Holt, kindly elucidated SNR
astrophysics.  Farhad Yusef-Zadeh and his collaborators kindly
provided us with an unpublished radio image, and the ACIS Hubble Deep
Field team gave us access to their data for calibration purposes. This
research was supported by NASA contract NAS 8-38252 and in part (SHP)
by JPL, under contract with NASA. Y.~M. was financially supported by
{\it the Japan Society for the Promotion of Science (JSPS)}.

\clearpage

\bibliography{apj-jour}

\clearpage

\figcaption[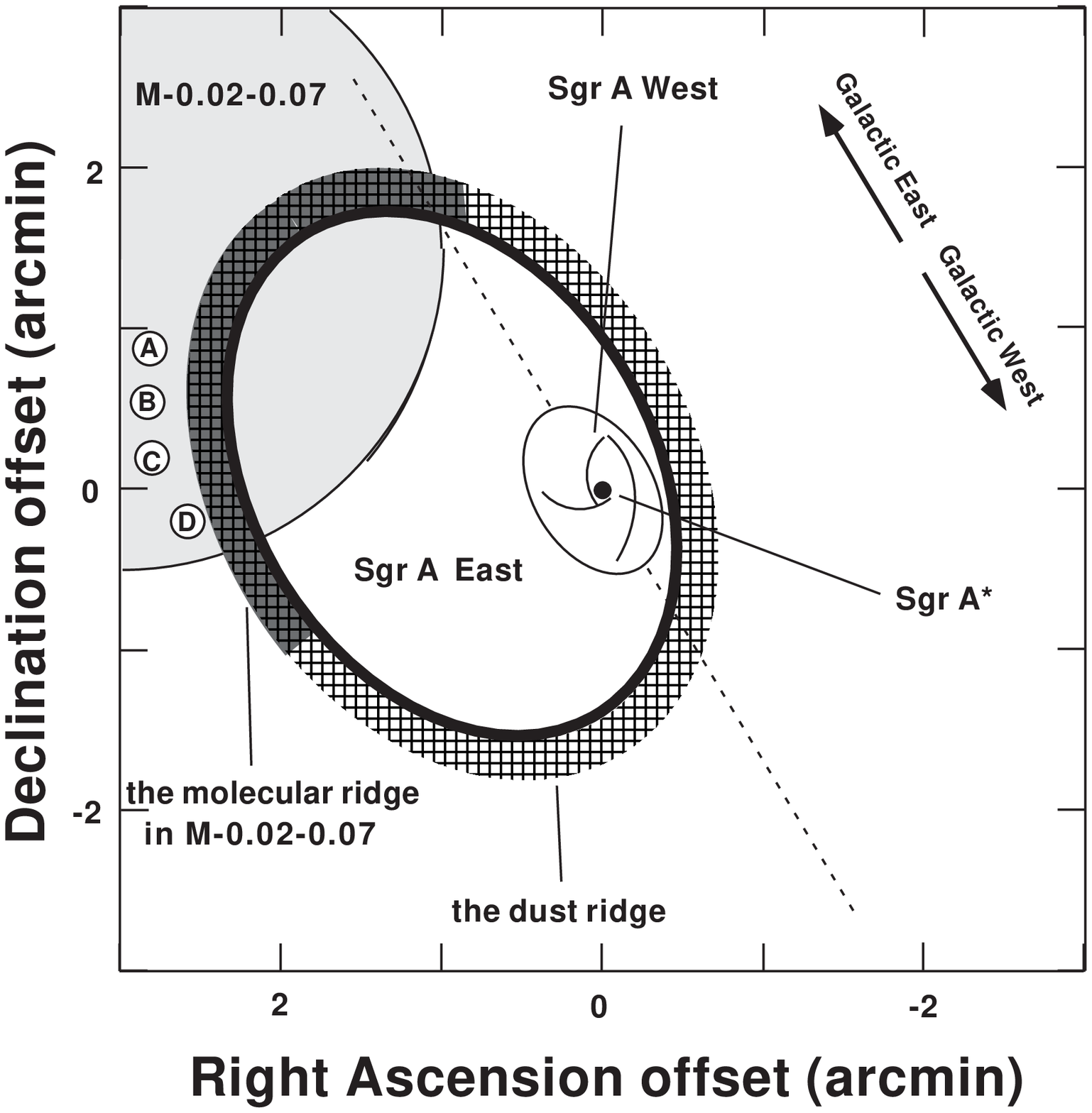]{
Schematic diagram showing the sky locations and rough sizes
and shapes of Galactic center sources \citep[e.g.,][]{Yusef00}. The
coordinate offsets are with respect to the compact non-thermal radio
source Sgr A* which coincides with the MBH. Sgr~A* is located at the
center of the thermal radio source Sgr~A West, which consists of a
spiral-shaped group of thermal gas filaments. 
Sgr~A West is surrounded by the molecular ring (also known as the
circumnuclear disk), the radius of which is about 30\arcsec.  The
non-thermal shell-like radio source Sgr~A East discussed here is
surrounding Sgr~A West, but its center is offset by about
$50\arcsec$. The non-thermal shell is surrounded by the dust and the
molecular ridge. The molecular cloud M$-0.02-0.07$ (the $+$50 km
s$^{-1}$ cloud) is located to the Galactic east of Sgr~A East. The
molecular ridge is also classified as a part of M$-0.02-0.07$ because
both of the peak velocities in molecular lines appear around $+$50
km~s$^{-1}$. At the eastern edge of the Sgr~A East shell, the chain of
HII regions (A-D) is seen. One arcminute corresponds to about 2.3 pc
at the distance of 8 kpc.
\label{figure:1}
}

\figcaption[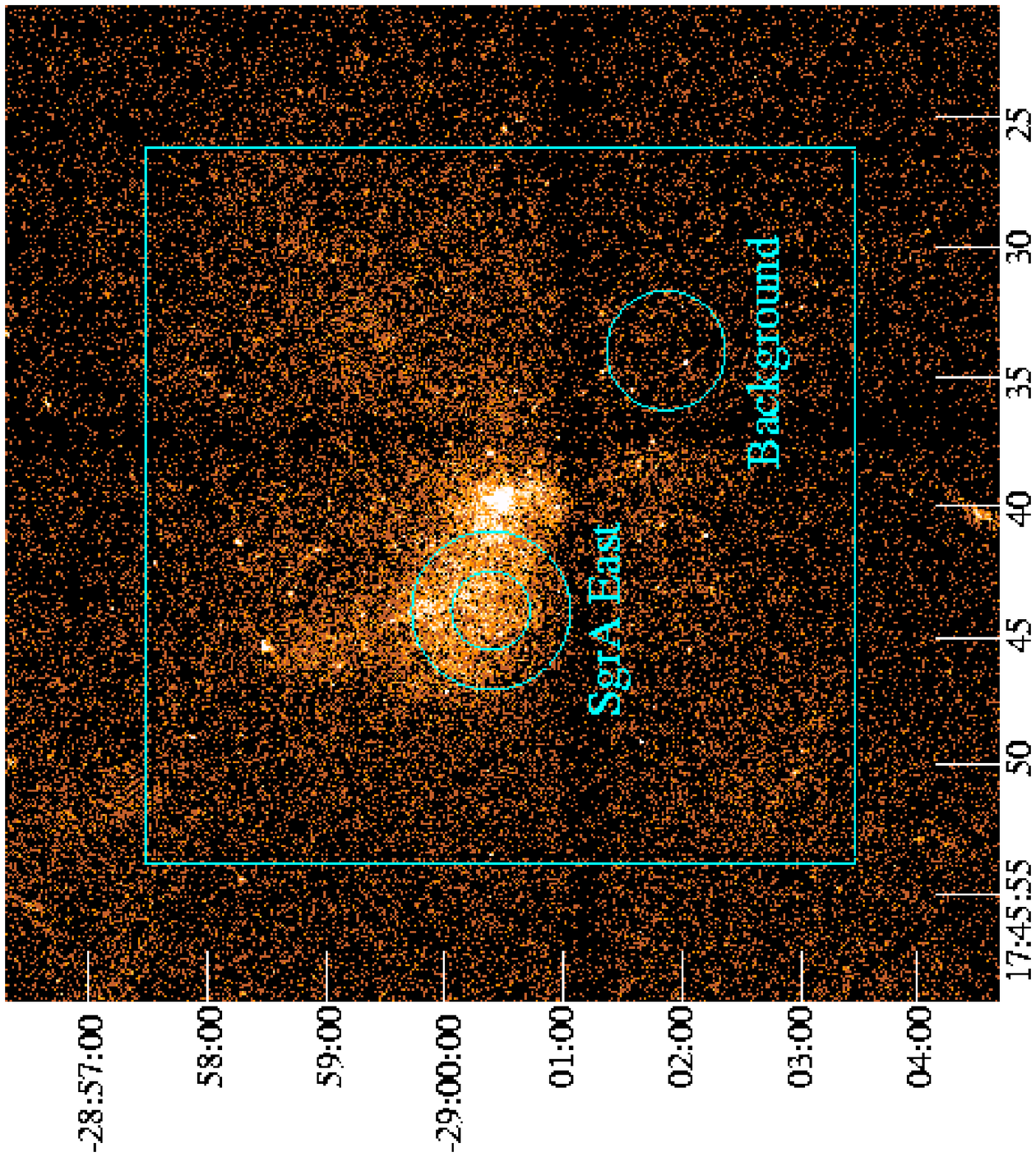,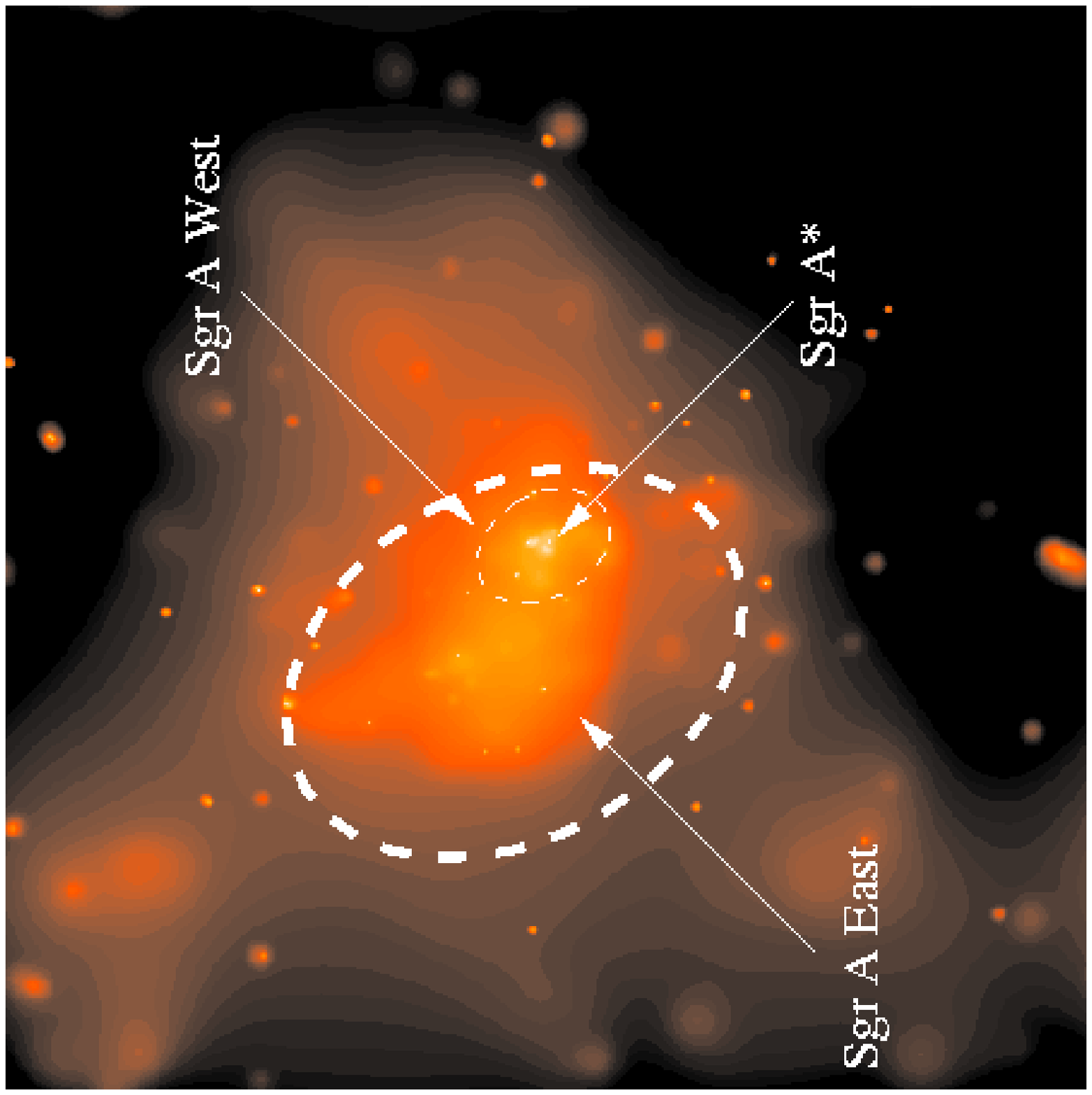,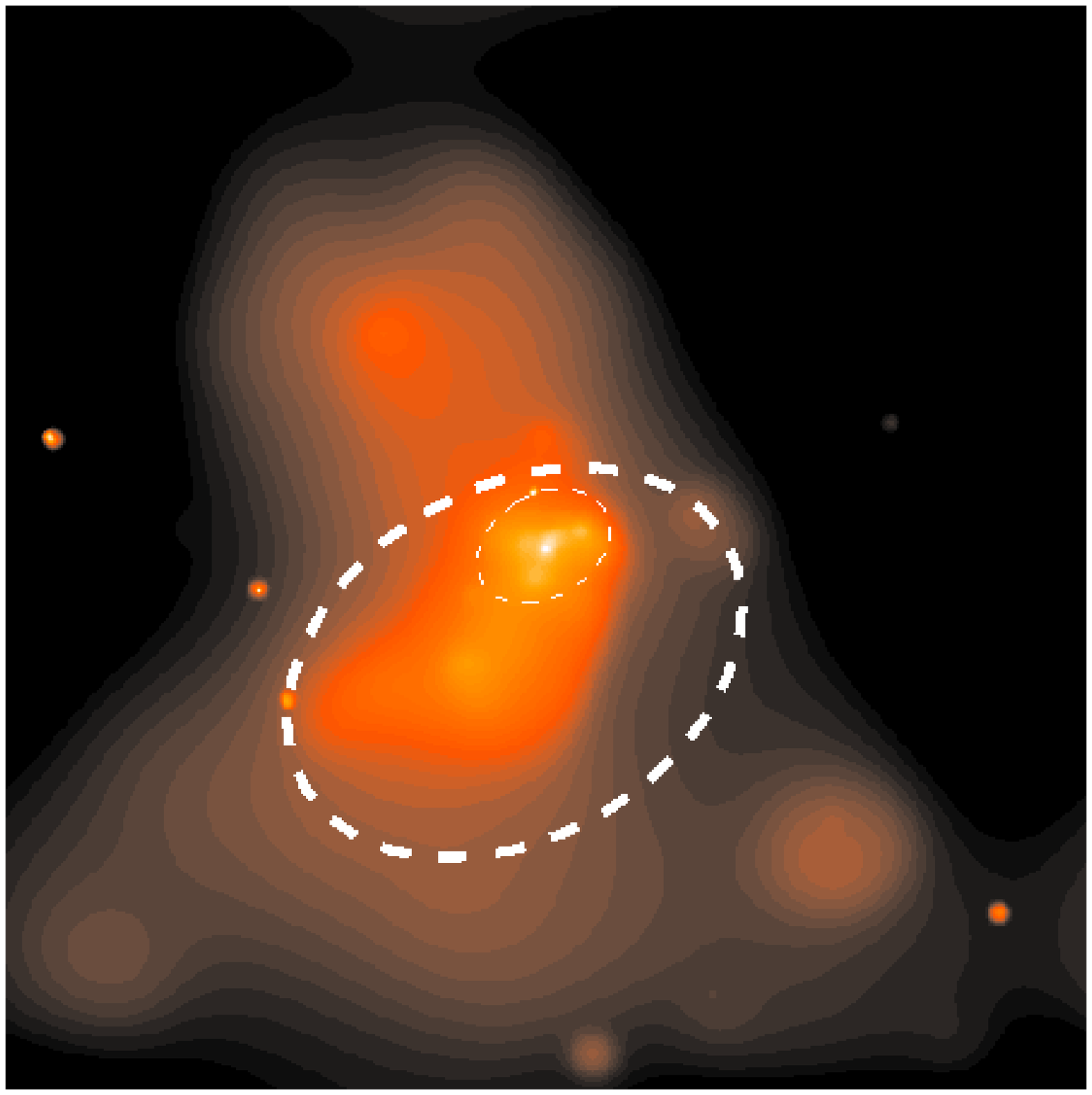,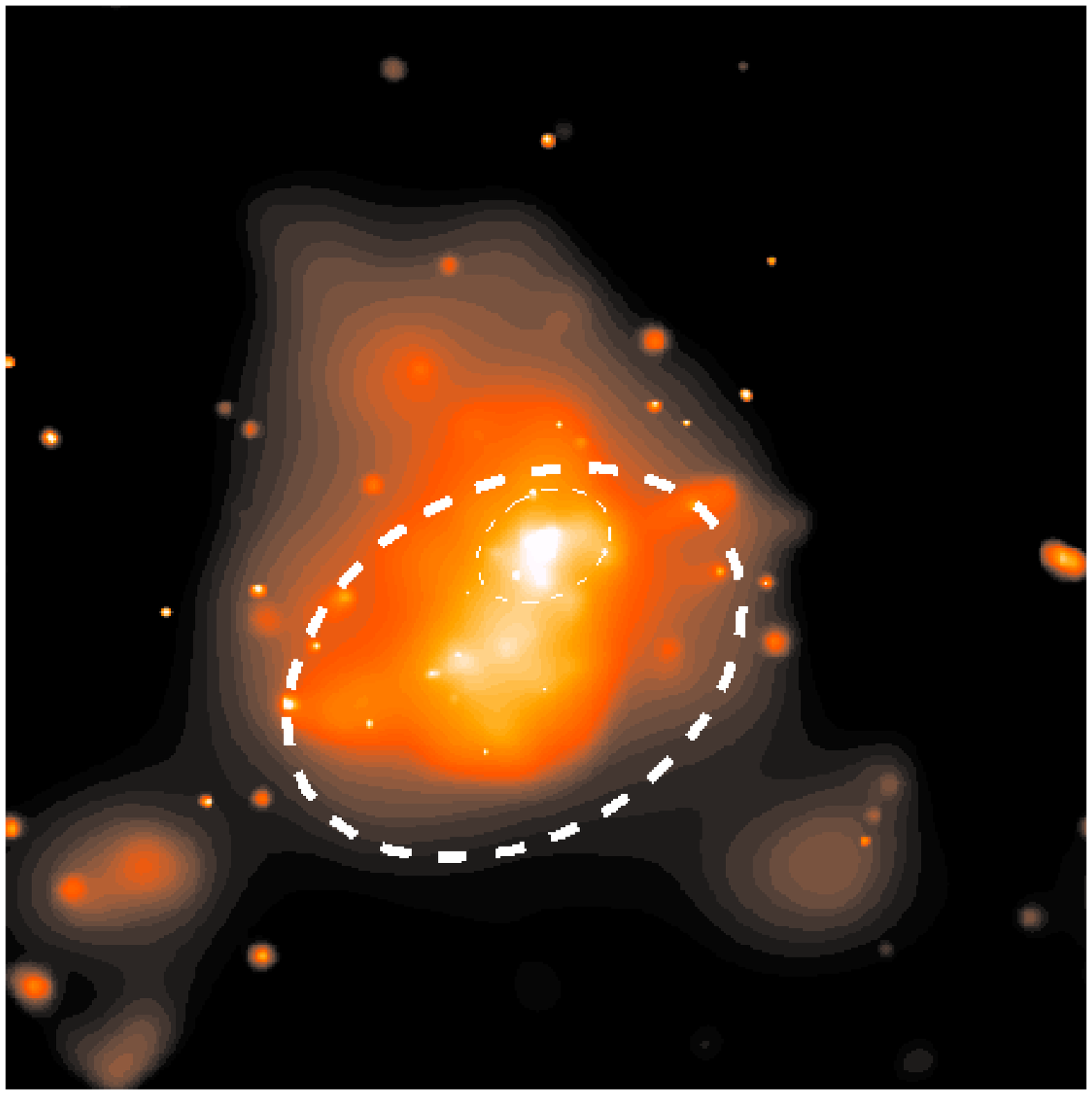,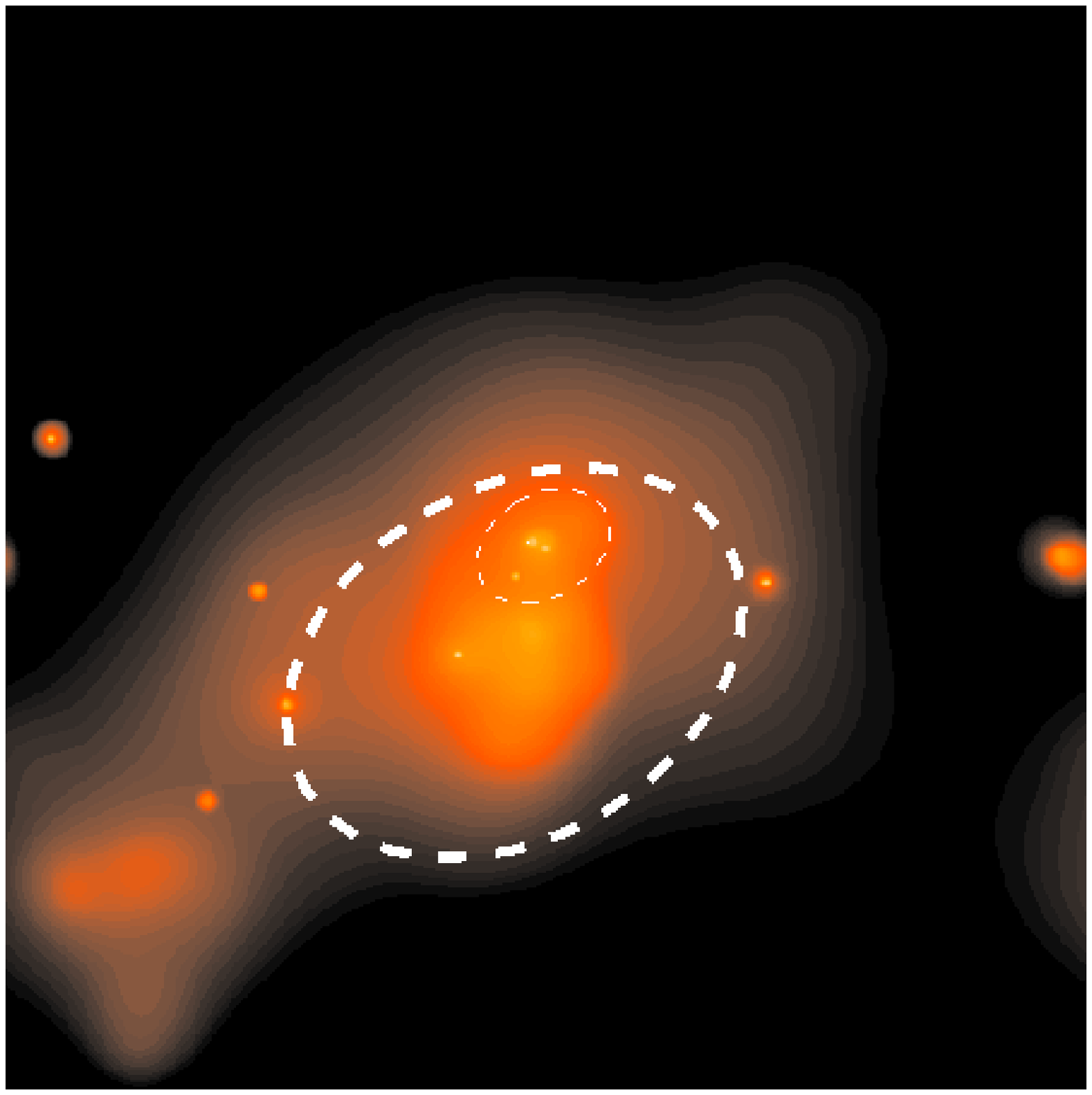]{
(a): Raw image in the 1.5--7.0 keV band binned by a factor of 2
pixels. 
The grids are the FK5 coordinates: RA(2000), Dec(2000).
Neither exposure nor vignetting correction are
applied. Circles correspond to the regions from which spectra
were taken. The outer circle of the Sgr~A East region is used for the
analysis of the overall spectrum, while the inner circle is used to
test radial dependence of the spectrum. The large square corresponds
to the scale of the schematic diagram given in
Figure~\ref{figure:1}. (b--e): Smoothed images in the 1.5--7.0 (b),
1.5--3.0 (c), 3.0--6.0 (d), and 6.0--7.0 (e) keV bands. Both exposure
and vignetting corrections were applied. The large and small white
dashed ellipses approximately represent the Sgr~A East non-thermal
shell and an outer boundary of the Sgr~A West region, respectively. In
all the panels, the field center is at Sgr~A*, and the panel size is
$8'.4\times8'.4$.
\label{figure:2}
}

\figcaption[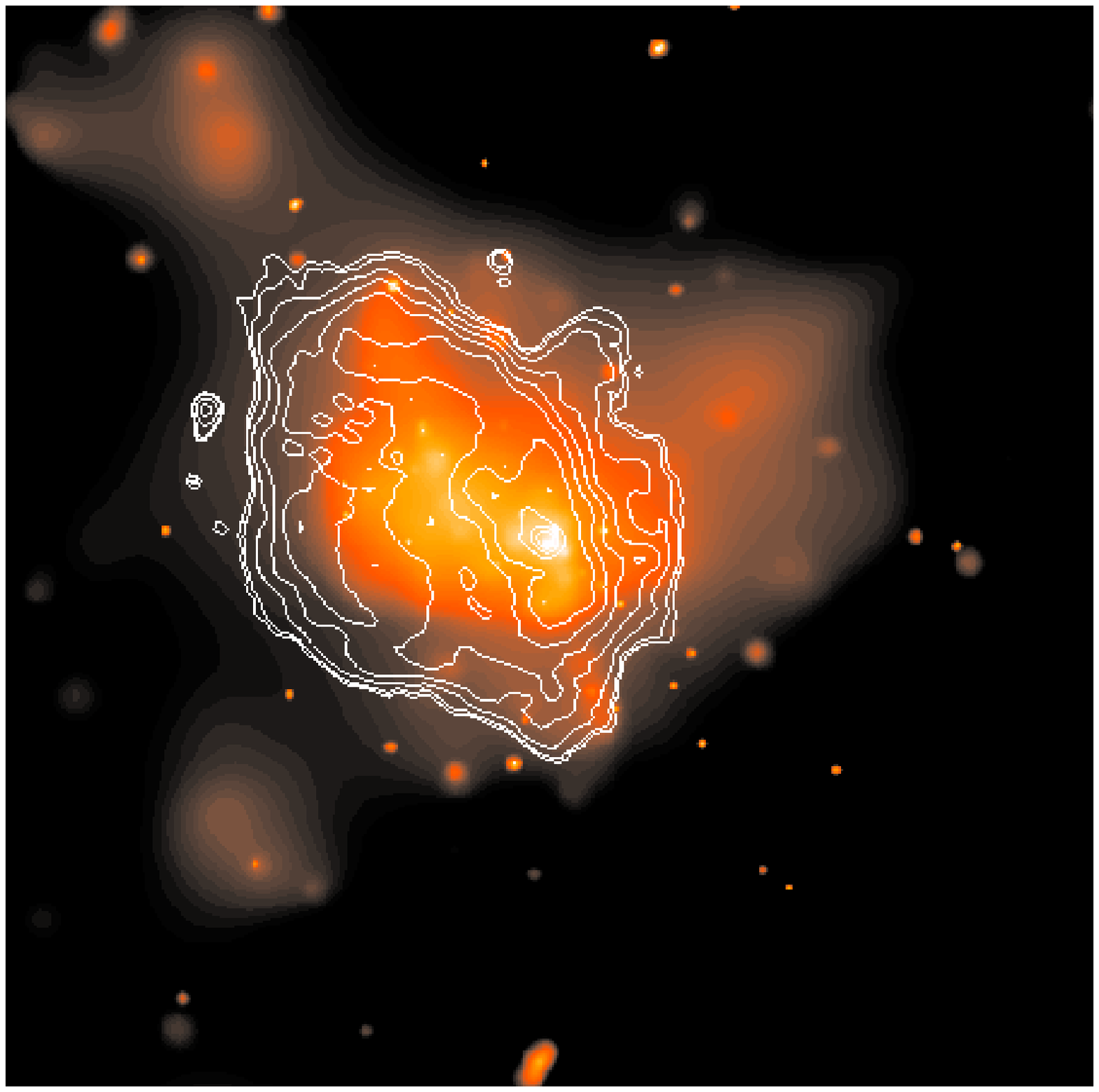]{
Smoothed X-ray image (1.5--7.0 keV) with 20~cm radio
contours (white: Yusef-Zadeh private communication). 
\label{figure:3}
}

\figcaption[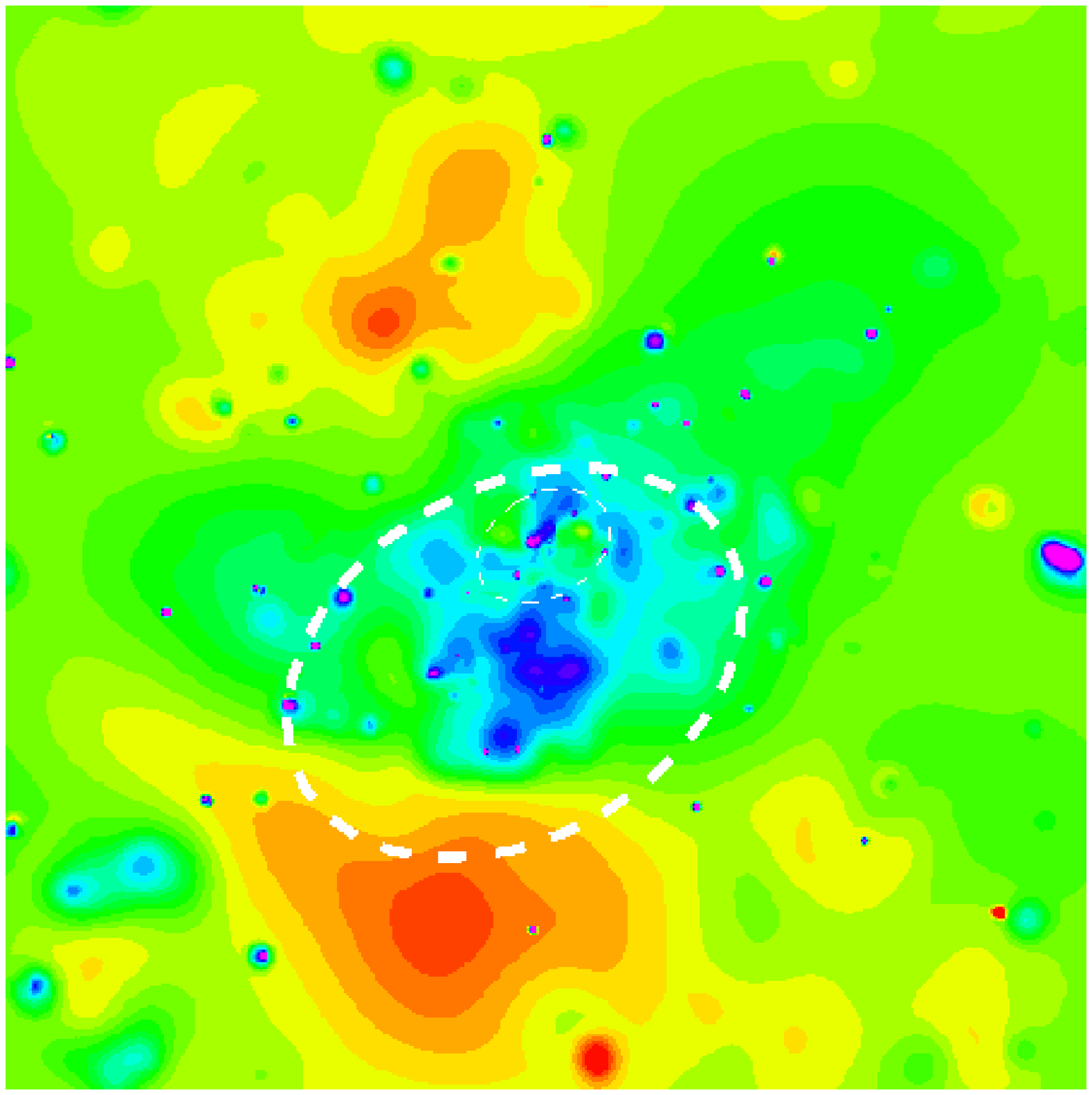,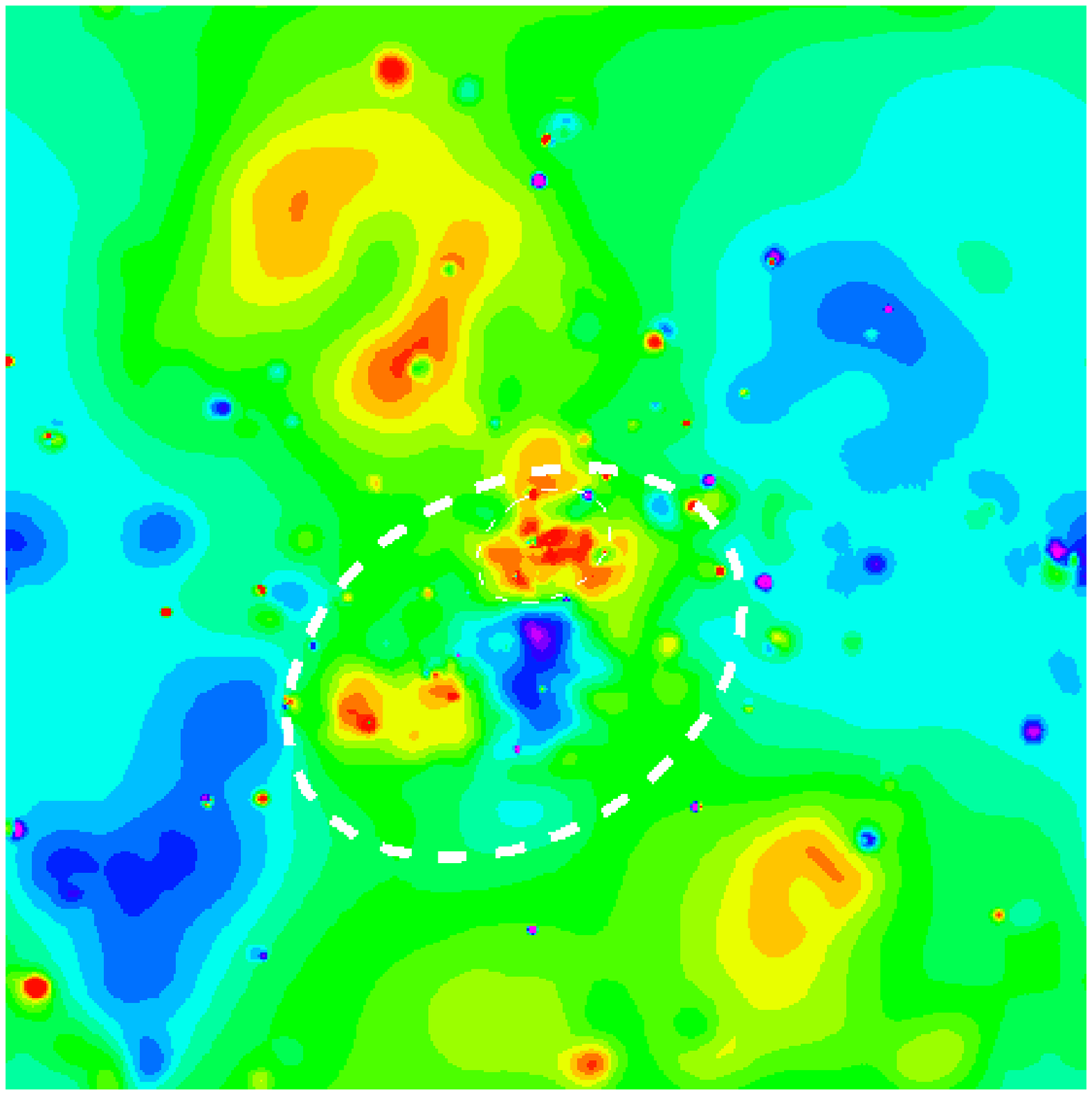]{
The hardness ratio map of the smoothed images: the blue
color means harder while the red implies softer. (a) (3.0--6.0~keV)/(1.5--3.0~keV). (b) (6.0--7.0~keV)/(3.0--6.0~keV). 
\label{figure:4}
}

\figcaption[figure5.ps]{
The ACIS-I spectrum of Sgr~A East. Error bars are
1 $\sigma$. The solid line corresponds to the best-fit value with the
MEKA model summarized in Table~2.  Fit residuals are shown in the bottom panel.
\label{figure:5}
}

\figcaption[figure6.ps]{
Spectra taken from two regions of Sgr~A East. The upper spectrum is
for the circular region with a radius of 20$^{\prime\prime}$ while the
lower is for the outer annular region with inner and outer radii
of 20$^{\prime\prime}$ and 40$^{\prime\prime}$. The unit in the
vertical axis is arbitrary.
\label{figure:6}
}

\figcaption[figure7a.ps,figure7b.ps]{
Plots of best fit parameters for two concentric annuli.
(a) Continuum. The units for $N_{\rm H}$ and $kT_e$ are [$10^{22}$
cm$^{-2}$] and [keV], respectively.  (b) Equivalent widths for
emission lines. The unit is [eV]. Note that the ordinates for both
plots on panel (b) are logarithmic and span one order of magnitude.
\label{figure:7} 
}

\figcaption[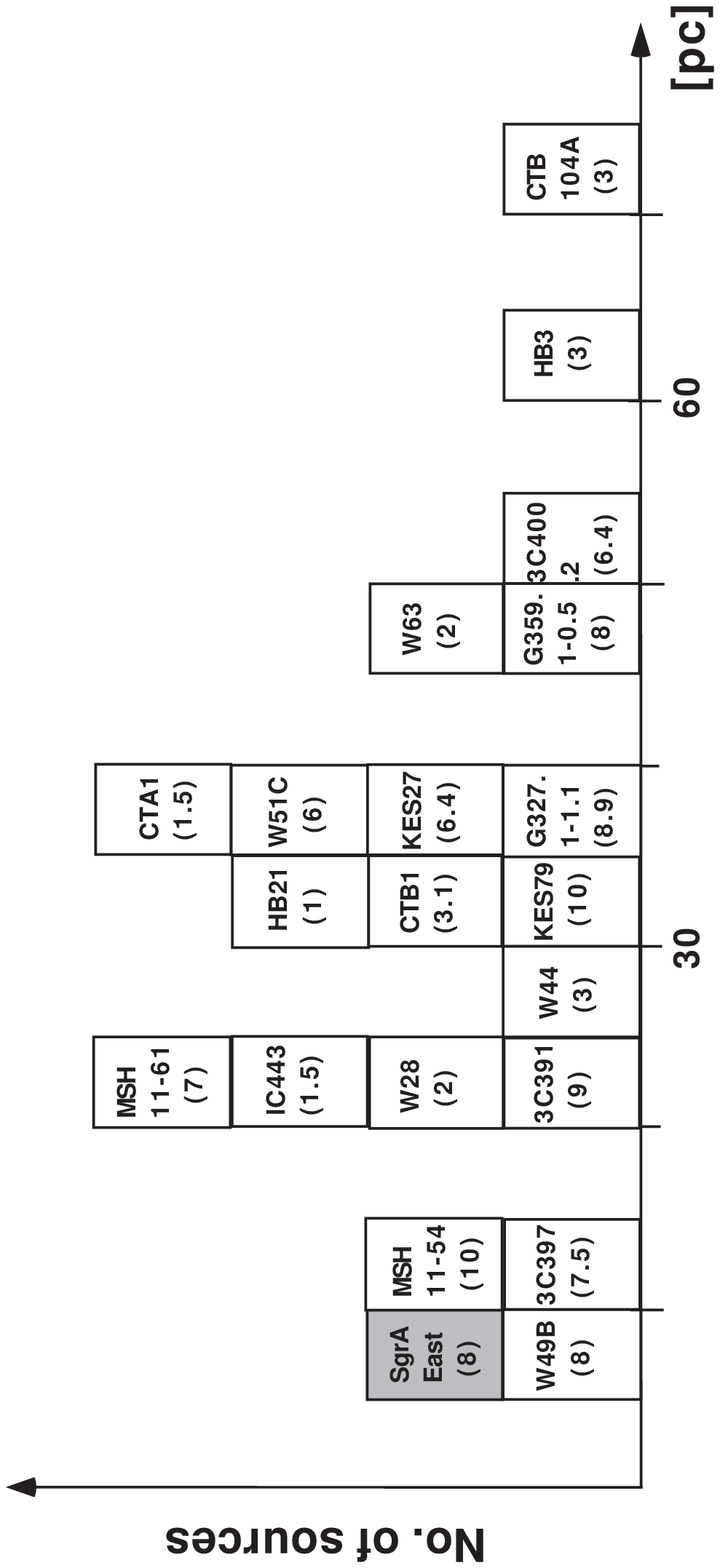]{
Histogram of the linear sizes of MM SNRs. Values given in
parentheses are the assumed distances in kpc \citep[see][and reference
therein]{Rho98, Bamba00}. 
\label{figure:8} 
}

\figcaption[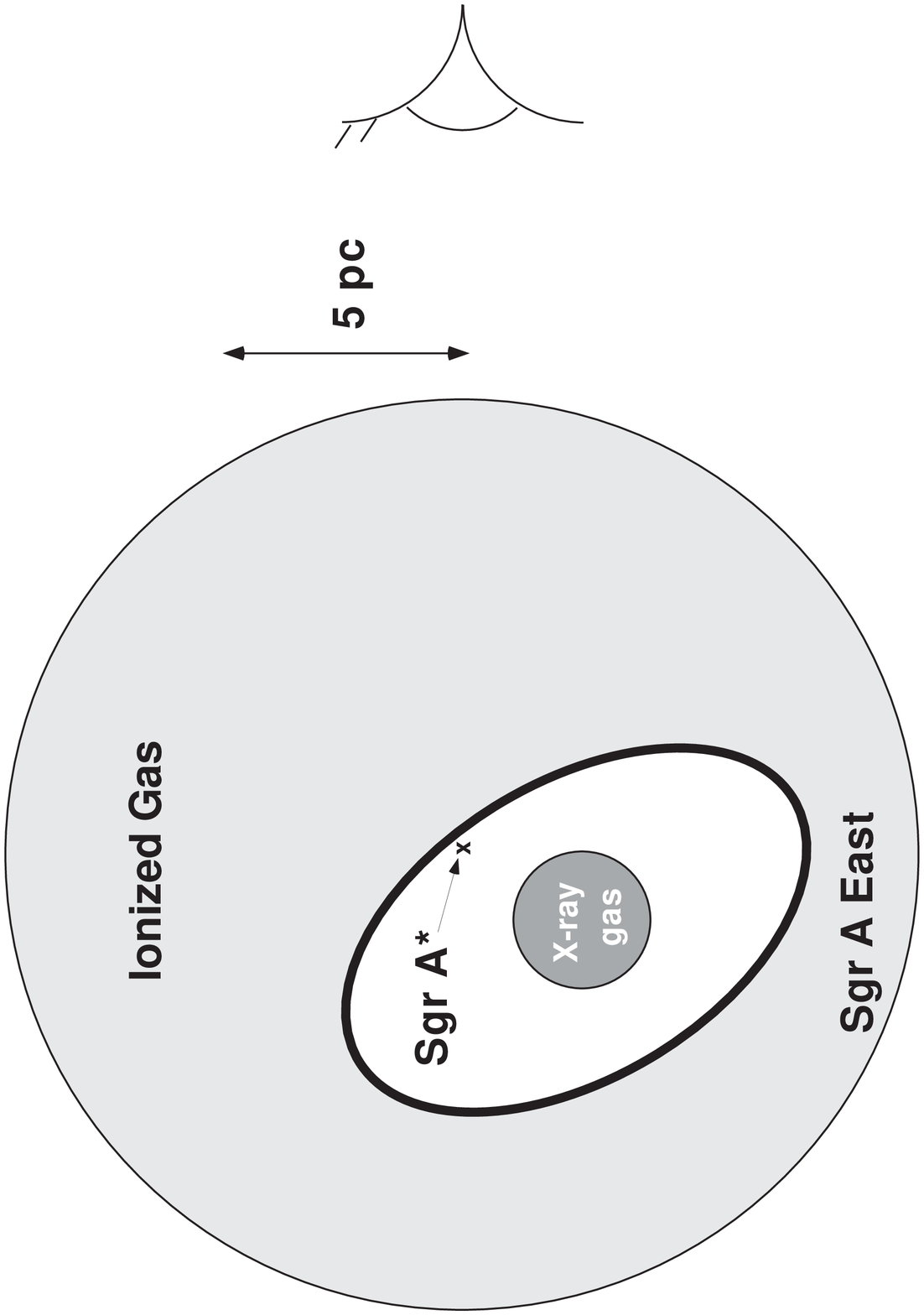]{
Schematic diagram of the relative positions and sizes of
Sgr~A*, Sgr~A East and the ionized gas halo along the line of sight from
the Sun with the positive Galactic longitude (east) being down. The
ionized gas halo of $10^3$ cm$^{-3}$ is rotating around Sgr~A* and is
filling the non-solid-body rotation region. A SNR, Sgr~A East, was
expanding into the ionized gas halo and the radio structure associated with
the slow forward shock was sheared by the Galactic rotation. The hot
ejecta plasma is centrally concentrated within the Sgr~A East radio
shell and is visible in X-rays. Sgr~A* was hit by the front edge of
the Sgr~A East shell in the recent past and is currently in the hot
cavity inside of the shell.
\label{figure:9}
}

\begin{deluxetable}{lcccc}
\tablewidth{0pt}
\tablecolumns{5}
\tablecaption{Best-fit parameters to the Sgr~A East spectrum fitted with
a thermal bremsstrahlung with four Gaussians \label{table:1}}
\startdata
\cutinhead{Continuum}
 &  \multicolumn{3}{l}{$N_{\rm H}$ [10$^{22}$ cm$^{-2}$]} & 9.4(8.7--10.2) \\
 & \multicolumn{3}{l}{$kT_e$ [keV]} & 3.0(2.6--3.5) \\  
\cutinhead{Emission lines}
  & Line energy [keV]         & \multicolumn{1}{l}{Line I.D. [keV]\tablenotemark{a}}       & $I$ [$10^{-5}$ ph cm$^{-2}$ s$^{-1}$]\tablenotemark{b}             & \colhead{$EW$ [eV]\tablenotemark{c}} \\
& 2.49(2.45--2.53)      & \multicolumn{1}{c}{S{\footnotesize XV} (2.45)}        & 0.9(0.5--1.4)             & 140         \\ 
& 3.16(3.10--3.23)      & \multicolumn{1}{c}{Ar{\footnotesize XVII} (3.14)}       & 1.1(0.7--1.6)             & 92         \\
& 3.85(3.81--3.88)      & \multicolumn{1}{c}{Ca{\footnotesize XIX} (3.90)}       & 2.2(1.7--2.6)             & 173         \\ 
& 6.69(6.67--6.71)      & \multicolumn{1}{c}{Fe{\footnotesize XXV} (6.67)}       & 12.2(11.2--13.2)          & 3100         \\ 
\cutinhead{2--10 keV band}  
& \multicolumn{3}{l}{$F_{\rm x}$ [erg cm$^{-2}$ s$^{-1}$]} & 5 $\times$ 10$^{-12}$         \\
& \multicolumn{3}{l}{$L_{\rm x}$ [erg s$^{-1}$]} & 8 $\times$ 10$^{34}$          \\ 
& \multicolumn{3}{l}{$\chi^2$(d.o.f.)} & 190(177)        \\ 

\enddata

\tablecomments{No systematic error was included in the values given in
the table. See the text for the error. X-ray flux ($F_{\rm x}$) is not
corrected for absorption while luminosity ($L_{\rm x}$) is
corrected. All the errors given in parentheses are for 90 \%
confidence level. A narrow line is assumed.}  
\tablenotetext{a}{Line identification. Theoretical energy of a
K$\alpha$ transition line (Mewe{,} Gronenschild \& van den Oord 1985)
is given in parenthesis.}  
\tablenotetext{b}{Line flux.}
\tablenotetext{c}{Equivalent width.}
\end{deluxetable}

\begin{deluxetable}{lc}
\tablewidth{0pt}
\tablecolumns{2}
\tablecaption{Best-fit parameters to the Sgr~A East spectrum fitted with
the MEKA model \label{table:2}}

\tablehead{
\colhead{Parameter [unit]  \hspace{2cm}   }                                 & \colhead{Best fit value}        
}

\startdata
$N_{\rm H}$ [10$^{22}$ cm$^{-2}$]& 11.4(10.5--12.3)        \\
$kT_e$ [keV]& 2.1(1.9--2.4) \\  
$Z$ & 3.9(2.9--5.9) \\  
Normalization & 1.1(0.9--1.3) \\  
$\chi^2$(d.o.f.)& 217(184)      \\
\enddata

\tablecomments{
Normalization: $10^{-12} \int_{}^{} n_{\rm e} n_{\rm H} dV$ / $(4
\pi D^2)$, where $n_{\rm e}$ is the electron number density
(cm$^{-3}$), $n_{\rm H}$ is the proton number density (cm$^{-3}$), and $D$
is the distance to the source (cm). $n_{\rm e}$ $=$ $1.8 \times
n_{\rm H}$ for the best fit.
}
\end{deluxetable}

\begin{deluxetable}{lcccc}
\tablewidth{0pt}
\tablecolumns{5}
\tablecaption{
Comparison between Sgr~A East and W49~B \label{table:3}
} 
\tablehead{
\colhead{} & \multicolumn{2}{c}{Observed quantities} & \multicolumn{2}{c}{References} \\
\colhead{} & \colhead{Sgr~A East} & \colhead{W49~B} & \colhead{Sgr~A East} & \colhead{W49~B} 
}
\startdata
Distance			& 8 kpc				& 8 kpc			& (1)		& (2,3)	\\ 
\cutinhead{Radio (non-thermal)}	
Morphology			& Shell (ellipse)		& Shell (roughly circular)& (4)	& (5)		\\		
Size				& $3'.5\times2'.5$		& $3'.5$		& (4)	& (5)		\\		
Spectral index ($\nu^\alpha$)	& $-0.8$			& $-0.5$			& (6)	& (6)	\\
Flux at 1 G Hz			& 100 Jy			& 38 Jy			& (6)	& (6)	\\ 
\cutinhead{X-rays (optically thin thermal)} 
Morphology			& Centrally concentrated			& Centrally concentrated		& this work		& (5)	\\
$kT_e$				& 2~keV				& 2~keV			& this work		& (7)	\\
$L_{\rm 2-6~keV}$		& $6\times10^{34}$ ergs s$^{-1}$ & $6\times10^{35}$ ergs s$^{-1}$	& this work		& (7) \\
Equivalent width of Fe K	& 3~keV				& 5~keV 		& this work		& (7)	\\
Intensity of Fe K		& $1\times10^{-4}$ ph cm$^{-2}$ s$^{-1}$		& $1\times10^{-3}$ ph cm$^{-2}$ s$^{-1}$ & this work		& (7)	  

\enddata

\tablecomments{(1) Reid 1993, (2) Radhakrishnan et~al. 1972, (3) Moffett \& Reynolds 1994, (4) Ekers et~al. 1975, (5) Pye et~al. 1984, (6) Green 1991, (7) Smith et~al. 1985.}

\end{deluxetable}

\end{document}